\documentclass[prx,reprint,preprintnumbers,nofootinbib,amsmath,amssymb,aps,floatfix]{revtex4-2}
\usepackage[utf8]{inputenc}
\usepackage [english] {babel} 
\usepackage{amsmath,amssymb,latexsym}
\usepackage{slashed}
\usepackage{graphicx}
\usepackage{epstopdf}
\usepackage{mathtools}
\usepackage{hhline}
\usepackage{multirow}
\usepackage[normalem]{ulem}

\usepackage{xcolor}
\usepackage{hyperref}

\def \sec{\begin{section}}
\def \esec{\end{section}}

\newtheorem{theorem}{Theorem}

\newtheorem{corollary}{Corollary}[theorem]

\newtheorem{conjecture}{Conjecture}

\renewcommand{\tilde}{\widetilde}
\renewcommand{\bar}{\overline}

\def \ep {\epsilon}

\def \Ec {\mathcal{E}}

\def \Tc {\mathcal{T}}

\def \Oc {\mathcal{O}}

\def \Hc {\mathcal{H}}

\def \Nc {\mathcal{N}}

\def \ra {\rightarrow}

\def \beq { \begin{equation}}
\def \eeq {\end{equation}}

\def\bal#1\eal{\begin{align}#1\end{align}}

\def \at {\biggl{\vert}}

\DeclareMathOperator*{\Tr}{Tr}
\def \id {\mathbf{1}}

\newcommand\AM[1]{{#1}}
\newcommand\AMnew[1]{{ #1}}

\renewcommand\Re{\operatorname{Re}}
\renewcommand\Im{\operatorname{Im}}

\let\ldash\l
\def \l {\left}
\def \r {\right}

\def \bra {\langle}
\def \ket {\rangle}

\begin{document}

\title{Observable and computable entanglement in time
}

\author{Alexey~Milekhin}
\author{Zofia~Adamska}
\author{John~Preskill}
\affiliation{Institute for Quantum Information and Matter, California Institute of Technology, Pasadena, CA 91125, USA}
\email{milekhin@caltech.edu}

\date{\today}

\begin{abstract}
    We propose a novel family of entanglement measures for time-separated subsystems. Our definitions are applicable to any quantum system, continuous or discrete. To illustrate their utility, we derive upper and lower bounds on time-separated correlation functions, akin to the bound on spatially separated correlators in terms of the mutual information. In certain cases our bounds are tight. 
    For relativistic quantum field theories our definition agrees with the analytic continuation from spacelike to timelike separated regions. We provide relevant measurement protocols and execute them on the IBM quantum device \texttt{ibm\_sherbrooke} for a simple qubit system. Also we perform explicit computations for an Ising spin chain, free fermions, (1+1)-dimensional conformal field theories and holographic theories.
     Finally we explain how the proposed entanglement in time provides a microscopic definition for the recently introduced timelike pseudoentropy. 

\end{abstract}

\maketitle

\section{Introduction}
\label{sec:intro}
Entanglement is a key property of quantum mechanics. Its applications range from classifying topological phases \cite{zeng2019quantum} to quantifying the black hole information paradox \cite{Almheiri:2020cfm}. However, entanglement is a property of a given \textit{state}; it is not sensitive to what happened to the system in the past. Generalizing the notion of entanglement to include information about time evolution could open a new window into probing the dynamics of quantum systems with many potential applications.

Guided by relativistic quantum field theory (RQFT), where space and time are intimately related, we propose a definition of \textit{``entanglement in time'' applicable to any quantum system, continuous or discrete.} Central to our construction is a generalization of the density matrix suited for characterizing the properties of timelike separated subsystems. We prove various properties of our ``timelike density matrix'' which are analogous to properties of the conventional ``spacelike'' density matrix.

To illustrate the utility of our construction, we use it to derive \textit{upper} and \textit{lower} bounds on dynamical (Wightman) correlation functions. We show that a certain measure of entanglement in time for two subsystems, which we call the \textit{entanglement imagitivity}, is nonzero if and only if one subsystem can influence the other at a later time. We also provide experimentally feasible protocols for measuring entanglement in time.

\section{Defining entanglement in time}
\label{sec:def}
\begin{figure}
\centering    
\includegraphics[scale=0.9]{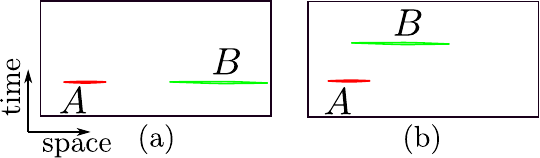}
    \caption{(a) Here $A,B$ are spacelike separated. There are many existing tools in RQFT to compute $\Tr \rho_{AB}^n$. (b) Timelike separated $A,B$. In RQFT this configuration can be continuously deformed to (a).} 
    \label{fig:AB}
\end{figure}

To motivate our construction, consider in an RQFT the von Neumann 
 entanglement entropy $-\Tr \rho_{AB} \log \rho_{AB}$ or Tsallis entropy $\Tr(\rho_{AB}^n )$ (exponentiated R\'enyi entropy) of two spatially separated regions $A$ and $B$; see Figure \ref{fig:AB} (a). These are analytic functions (possibly with poles and branch-cuts) of spacetime coordinates defining $A$ and $B$, which are explicitly known in some cases.
Now we can ask: if we analytically continue the answer for $\Tr(\rho_{AB}^n )$ from spacelike separation to timelike separation as in Figure \ref{fig:AB}(b), can we give it any quantum-mechanical interpretation?  For example, in a (1+1)-dimensional conformal field theory (CFT), the aforementioned analytic continuation simply means computing time-separated correlation functions of twist operators.
Can we still represent the answer as a trace over a certain Hilbert space, i.e. $\Tr T_{AB}^n$, where $T_{AB}$ is an analogue of the density matrix that can be defined when $A$ and $B$ are \textit{timelike} separated regions? Note that $A$ and $B$ do not need to be the same size or related to each other in any particular way.

To construct $T_{AB}$ for any system, not necessarily an RQFT, and to provide a useful perspective on it, recall the relation between density matrices and correlation functions. 
A basic rule of quantum mechanics is that the expectation value of observable $\Oc$ in a state with density matrix $\rho$ is given by
\beq
\bra \Oc \ket = \Tr (\Oc \rho).
\eeq
Suppose we bipartition our quantum system into a subsystem $A$ and its complement $\bar{A}$.
We can represent an expectation value of an operator $\Oc_A$ supported on $A$ as the contraction
\beq
\includegraphics{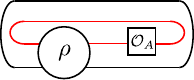}
\eeq
where $A$ is marked in red.
If we want to consider all possible operators $\Oc_A$ it is convenient to ``break'' the tensor diagram where $\Oc_{A}$ is inserted, and leave the corresponding ``slot'' open \cite{Calabrese:2004eu,Narayan:2023ebn}:
\beq
\includegraphics{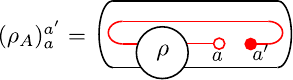}
\eeq
This logic naturally leads us to the reduced density matrix $\rho_A$ on $A$. We use open and filled circles in tensor diagrams to indicate input indices (belonging to the dual Hilbert space) and output indices (belonging to the original physical Hilbert space).

Let us repeat the same logic, but for the \textit{dynamical} (Wightman) correlation function:
\beq
\label{eq:wightman}
\bra \Oc_A(0) \Oc_B(t) \ket = 
\Tr( \rho \Oc_A U^\dagger \Oc_B U),
\eeq
where $U$ may be interpreted as a unitary time evolution operator.
Now $A$ and $B$ represent parts of the system at \textit{different times}. They do not have to be related to each other in any particular way.
By representing the contraction (\ref{eq:wightman}) as a tensor diagram and breaking it where $\Oc_A$ and $\Oc_B$ are inserted, we arrive at the \textit{spacetime} density matrix $T_{AB}$:
\begin{center}
\beq
\label{eq:TAB}
\includegraphics{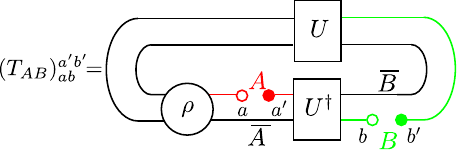}
\eeq
\end{center}
$T_{AB}$ is a natural analogue of the conventional density matrix. 
It is straightforward to generalize this definition to an arbitrary quantum channel $\Ec$, by replacing $U,U^\dagger$ with the appropriate sum of Kraus operators.

If $A$ and $B$ constitute the entire system before and after the evolution respectively, one can write $T_{AB}$ as a matrix product  
\beq
\label{eq:alg}
T_{AB} = J \ (\rho \otimes \id_B) ,
\eeq
where
\beq
J = \sum_{i,j=1}^{\dim \Hc} \Ec(| i \ket \bra j |) \otimes | j \ket \bra i |.
\eeq
The state $J$ is sometimes called the Jamio{\ldash}kowski state associated with the quantum channel $\mathcal{E}$. (It is related to the Choi state by a partial transposition.)

Exactly like a ``normal'' density matrix on $AB$, $T_{AB}$ is a linear map from $\Hc_A \otimes \Hc_B$ to itself:
\beq
T_{AB}: \Hc_A \otimes \Hc_B \ra \Hc_A \otimes \Hc_B.
\label{eq:h_map}
\eeq
However, notice that here $\Hc_A$ and $\Hc_B$ are completely \textit{independent} linear spaces; they are \textit{not} necessarily related by Schr\"odinger evolution. Correspondingly, operators $\Oc_A$ on $A$ and $\Oc_B$ on $B$ are not related by Heisenberg evolution and in fact they commute.
Their contractions with $T_{AB}$ produce Wightman correlation functions:
\beq
\Tr(T_{AB} (\Oc_A \otimes \id_B) (\id_A \otimes \Oc_B)) = \bra \Oc_A(0) \Oc_B(t) \ket.
\eeq
This property can be taken as the definition of $T_{AB}$.
The operator ordering on the right-hand side is hard-coded into the definition of $T_{AB}$. 
How, then, can the opposite operator ordering be represented?

One can easily check that $T^\dagger_{AB}$ is given by
\beq
    \includegraphics{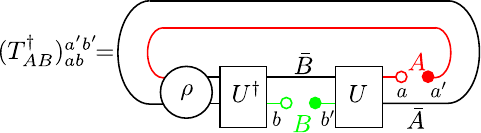}
\eeq
and hence contractions with it yield the opposite ordering:
\beq
\Tr(T^\dagger_{AB} \Oc_A \otimes \Oc_B) = \bra  \Oc_B(t) \Oc_A(0) \ket.
\eeq
Thus we see that $T_{AB}$, in contrast to the standard density matrix, need not be hermitian.  This property arises because operator ordering can be important when operator insertions are separated in time 
--- a feature absent for nonoverlapping regions lying on the same time slice in a theory respecting relativistic causality.

Going back to our original motivation, we can now show that, in RQFT, $\Tr T_{AB}^n$ coincides with the analytic continuation of $\Tr \rho^n_{AB}$.
In RQFT we can represent a density matrix by introducing 
\cite{Calabrese:2004eu} spacetime cuts at the locations of $A,B$ where the fields can take different values on the two sides of a cut.
For example, with Euclidean state preparation, the corresponding path integral looks like (crosses indicate possible source insertions):
\beq
    \includegraphics{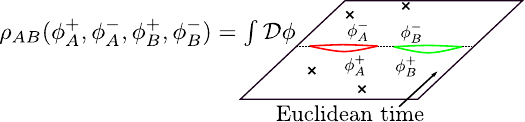}
\eeq
Traces of $\rho_{AB}^n$ are evaluated by taking several such manifolds together and gluing them cyclically along $A$ and $B$ to reproduce matrix multiplication. Moving region $B$ in time means that the boundary conditions are evolved under Heisenberg evolution:
\beq
    \includegraphics{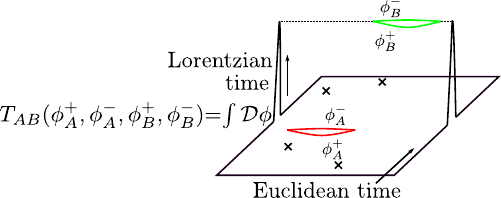}
\eeq
This diagram is exactly $T_{AB}$ in (\ref{eq:TAB}), where input and output indices $a,a'$ play the role of the field values. The only thing to keep in mind is that $B$ always follows $A$ in the operator ordering (or vice versa), which is enforced by imposing the appropriate RQFT $i \varepsilon$ prescription.

\section{Relation to other approaches}
In Appendix \ref{app:comp}, we briefly review other proposals for describing ``entanglement in time'' and explain how they differ from ours. The main conceptual difference is that in most previous proposals both forward and backward time evolution are considered, while in contrast we consider only forward time evolution. Put another way, in other proposals operators can be inserted on both the forward and backward branches of the Keldysh contour, but we allow insertions on only the forward branch. For this reason, our formalism is well suited for studying time correlations of observables, while other formalisms are more appropriate for studying the effects of measurements.

Our spacetime correlation tensor $T_{AB}$ can be viewed as a kind of Kirkwood-Dirac distribution \cite{Arvidsson-Shukur:2024fhs} \AMnew{or the "left bloom" of \cite{Fullwood:2022rjd}} specifically constructed so that Wightman correlation functions are obtained by contracting observables with $T_{AB}$. Its construction is simple, allowing us to derive general analytic results, and relate these results to explicit computations in RQFT and holographic theories. Furthermore, $T_{AB}$ itself can be directly measured.

Previous proposals, like ours, aimed to generalize the density matrix to capture correlations in time as well as space. For example, Ref. \cite{Cotler:2017anu} introduced a ``superdensity matrix'' to characterize dynamics. The superdensity matrix, unlike our $T_{AB}$, is hermitian, but it is not convenient for analyzing time-dependent correlation functions and its moments are not related to entropies in RQFTs.
Related constructions have been explored in the literature before under the names quantum comb \cite{Chiribella:2007igb}, pseudo-density matrix \cite{Fitzsimons:2013gga}, and Feynman--Vernon functional \cite{Feynman:1963fq,sonner2021influence,Carignano:2023xbz,Carignano:2024jxb,Bou-Comas:2024pxf}.
More recently, \cite{Glorioso:2024xan} introduced a notion of mutual information in time which can bound both commutators and anticommutators of time-dependent observables. 
However, their definition is not explicit and involves a certain extremization procedure. 
\AMnew{Another notion of mutual information in time can bound the amount of information possible to extract from sequential measurements \cite{Wu:2024ucp}.}
None of these approaches are related to the analytic continuation in RQFT, except \cite{Doi:2023zaf,Carignano:2024jxb} which studied entropic quantities for a single \textit{time-like} region. In certain cases this can be related to our construction as we will explain in Sec. \ref{subsec:pseudoentropy}.

We stress again that $T_{AB}$, though inspired by RQFT, \textit{is explicitly defined for any quantum system, continuous or discrete, and whether or not the dynamics is Lorentz invariant}. In this paper we will explore two types of entanglement in time: 
Tsallis entropies $\Tr T_{AB}^n$ and a novel notion of \textit{entanglement $p$-imagitivity} $||T_{AB} - T_{AB}^\dagger||_p$, where $\|\cdot\|_p$ denotes the Schatten $p$-norm.

For RQFTs, Tsallis entropies of $T_{AB}$ can be obtained from analytic continuation from spacelike separated regions:
\beq
\text{RQFT:} \ \Tr T_{AB}^n = \Tr \rho_{AB}^n \at_\text{analytic continuation}
\label{eq:an_cont}
\eeq
In other words, using the RQFT terminology, these quantities are obtained from correlators of time-separated twist-operators.
We will illustrate this by performing explicit computations in (1+1)-dimensional conformal field theories and holographic theories.
Taking the limit $n \ra 1$ to obtain von Neumann entropy is subtle and we will address it later. We will explain that in some cases our notion of entanglement in time coincides with the recently introduced timelike pseudoentropy.

Entanglement $p$-imagitivities $||T_{AB} - T_{AB}^\dagger||_p$ cannot be obtained from a simple analytic continuation from $\Tr \rho_{AB}^n$.
We will demonstrate their usefulness by using them to derive upper and lower bounds on time-separated correlation functions.

We call $\Tr T_{AB}^n$ and $||T_{AB}-T_{AB}^\dagger||_p$ \textit{entanglement} entropies in accord with the standard terminology in which $\Tr \rho_{AB}^n$ is called the Tsallis (or R\'enyi) entanglement entropy. However, for multipartite systems the Tsallis entropy does not necessarily quantify coherent quantum correlations; it could signal classical correlations instead. The same caveat applies to $\Tr T_{AB}^n$. Moreover, in the timelike setting there is no standard definition of entanglement and thus in this paper we will speak of ``entanglement entropy in time'' without offering a sharp operational interpretation of the ``entanglement.''
However, we will show that nonzero 2-imagitivity $||T_{AB}-T_{AB}^\dagger||_2$ 
is necessary and sufficient (\textbf{Theorems} \ref{th1}, \ref{th2}), and nonzero $\Im \Tr T_{AB}^2$ or negative $\Re \Tr T_{AB}^2$ is sufficient (\textbf{Corollary} \ref{col:im_bound}), for time-dependent commutators $[\Oc_A(0), \Oc_B(t)]$ to have non-vanishing expectation values, \AMnew{which is equivalent to signaling.} 
\AMnew{It would be interesting to investigate how this notion of entanglement in time is related to the violation of temporal Bell inequalities \cite{Brukner:2004egd,leggett1985quantum,Emary_2013,zych2019bell}.}

\section{General properties}
\label{sec:properties}
The spacetime density matrix defined above has very nice properties:

1. For RQFT the R\'enyi entropies of $T_{AB}$ coincide with the answer one obtains via the analytic continuation from spacelike separated $AB$ as in eq.~(\ref{eq:an_cont}).

2. Tracing out $A$ (or $B$) yields the conventional density matrix $\rho_B$ (or $\rho_A$):
   \begin{align}
    &\Tr_B T_{AB} = \Tr_{\bar A} \rho=\rho_A, \nonumber\\
    &\Tr_A T_{AB} = \Tr_{\bar B} U\rho U^\dagger= \rho_B.
    \end{align}

    3. As a corollary, $T_{AB}$ has unit trace:
    \beq
    \Tr T_{AB} = 1.
    \eeq

    4. More generally, the analogue of $\Tr \rho^n \le 1$ for $n \ge 2$ is
    \beq
    |\Tr T_{AB}^n| \le 1.
    \eeq
    We will prove this inequality when we discuss the measurement protocols in Sec.~\ref{sec:measure}.

    5.  In general $T_{AB}$ is not hermitian and does not commute with its adjoint
    \beq
    T_{AB} \neq T_{AB}^\dagger, \  [T_{AB}, T_{AB}^\dagger] \neq 0.
    \eeq

6. It is a standard result for pure states that the reduced density matrix on $A$ and the reduced density matrix on the complement $\bar{A}$ have the same spectrum. An analogous property holds for $T_{AB}$.
For a pure state $\rho$, the spectrum is unchanged if one replaces $A$ by $\bar A$ and takes the partial transpose; that is, $T_{AB}$ given by eq.~(\ref{eq:TAB}) has the same spectrum as $T_{\bar{A}^t B}$ given by
\begin{center}
\beq
\label{eq:tildeTAB}
\includegraphics{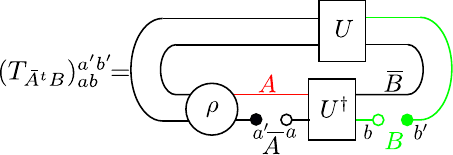}
\eeq
\end{center}
We prove this in App.~\ref{app:spectrum}. 
As a corollary, $T_{AB}$ and $T_{\bar{A} \bar{B}}$ have the same spectrum, again assuming that $\rho$ is pure.

7. Conventional density matrices have eigenvalues no larger than one. The corresponding property for $T_{AB}$ is that all its singular values are no larger
than $1$: 
\beq
||T_{AB}||_\infty \le 1.
\eeq
This implies that all eigenvalues lie within the unit disk. 
We prove this in App.~\ref{app:bound-values}.

8. $T_{AB}$ can be used to bound correlation functions by applying H\"older's inequality\footnote{The Schatten $p$-norm of a matrix $M$ is defined as
$$
||M||_p = \l( \Tr (M^\dagger M)^{p/2} \r)^{1/p}.
$$
}
\beq
|\Tr M N | \le \|M\|_p \|N\|_q,\quad 1/p+1/q =1.
\eeq
We find
\beq
\label{eq:c_bound}
\begin{split}
|\bra \Oc_A(0) \Oc_B(t) \ket|=
|\Tr( T_{AB} \Oc_A \otimes \Oc_B)| \\
\le || T_{AB}||_p ||\Oc_A||_q ||\Oc_B||_q, \ 1/p+1/q=1.
\end{split}
\eeq

Similarly, we can also bound the commutator/anti-commutator, e.g.:
\beq
\label{eq:bound}
\begin{split}
|\bra [\Oc_A(0), \Oc_B(t)] \ket|= \\
=|\Tr( (T_{AB}-T_{AB}^\dagger ) \Oc_A \otimes \Oc_B)| \le \\
\le || T_{AB}-T_{AB}^\dagger||_p ||\Oc_A||_q ||\Oc_B||_q, \\
1/p+1/q=1.
\end{split}
\eeq
Note that these bounds are not tight for the following reason. Mathematically, H\"older's inequality is tight, but the inequality might be saturated for a complicated operator $\Oc_{AB}$ which is not factorizable into $\Oc_A \otimes \Oc_B$. In Sec.~\ref{sec:bounds-commutators} we will show how to circumvent this problem.

9. Unfortunately, it is not obvious how to formulate subadditivity and strong subadditivity because $\Tr T_{AB}^n$ is complex in general.
Audenaert has proven \cite{Audenaert_2007} subadditivity of $p>1$ R\'enyi entropies in the following form:

\begin{theorem} (Audenaert) For any density matrix $\rho_{AB}$ on a finite-dimensional Hilbert space $\Hc_A \otimes \Hc_B$ and $p>1$:
\beq
1+||\rho_{AB}||_p \ge ||\Tr_A \rho_{AB}||_p + ||\Tr_B \rho_{AB}||_p.
\eeq
\end{theorem}
We have obtained some numerical evidence that $T_{AB}$ satisfies the same inequality, so we will formulate it as a conjecture:
\begin{conjecture}
    For any $T_{AB}$ on a finite-dimensional $\Hc_A \otimes \Hc_B$ and $p>1$:
    \beq
1+||T_{AB}||_p \ge ||\Tr_A T_{AB}||_p + ||\Tr_B T_{AB}||_p.
\eeq
\end{conjecture}

In App. \ref{app:mono} we prove a special case of this statement:
\begin{theorem}
\label{th:mono}
    Suppose that the initial state $\rho=|\psi_A \ket \bra \psi_A | \otimes \rho_{\bar{A}}$. Then for any $p>0$, 
    \beq
    ||T_{AB}||_p \ge ||\Tr _A T_{AB}||_p = ||\rho_B||_p.
    \eeq
\end{theorem}
For other non-hermitian analogues of density matrices, $||\cdot||^p_p$ has been referred to in the literature as the SVD R\'enyi entropy \cite{Parzygnat:2023avh,Caputa:2024qkk}.

We will finish this section with a comment regarding von Neumann entropy. In general, the computation of the von Neumann entropy  $-\Tr T_{AB} \log T_{AB}$ (and the related mutual information) is ill-posed for generic systems because $T_{AB}$ has complex eigenvalues, so the logarithm is not well-defined. However, in systems with continuous time dynamics (even if not relativistic) there is a way to make sense of this expression by starting from the conventional entanglement entropy when $A$ and $B$ lie on the same time slice and then continuously moving $B$ forward in time while keeping track of the phases in $\log T_{AB}$. 

We note that $\Tr T_{AB}^n$ and $-\Tr T_{AB} \log T_{AB}$ have UV divergencies in RQFT. This is true whether $A$ and $B$ have spacelike or timelike separation. Given that these quantities can be computed from the four-point correlation functions of twist operators (as is also the case for entropies of conventional density matrices), we expect these divergencies can be subtracted in the same way whether the separation of $A$ and $B$ is spacelike or timelike. In particular, we conjecture:
\begin{conjecture}
The analogue of mutual information
\beq
-\Tr \rho_A 
\log \rho_A - \Tr \rho_B \log \rho_B
+ \Tr T_{AB} \log T_{AB},
\eeq
is UV-finite in RQFTs.
\end{conjecture}

Unfortunately, $\Tr(T_{AB} T_{A B}^\dagger)$ is not always well-defined in the continuum limit. As we explain in App. \ref{app:ttbar}, it is very sensitive to lattice effects. In certain kinematical regimes, the continuum RQFT description predicts $T_{AB}=T_{AB}^\dagger$ but the lattice value of $\Tr(T_{AB} T_{AB}^\dagger)$ can be much larger than $\Tr(T_{AB}^2)$ by a factor $\sim e^{1/a}$ which is exponentially large in the lattice spacing $a$.

\section{Bounds on commutators and the Lieb--Robinson bound}
\label{sec:bounds-commutators}
Equation (\ref{eq:bound}) provides an upper bound on the commutator $\bra [\Oc_A(0), \Oc_B(t)] \ket$ in terms of the matrix norm $||T_{AB}-T^\dagger_{AB}||_p$, which we dub \textit{entanglement $p$-imagitivity} (the name mimics entanglement negativity \cite{Vidal:2002zz}), because it is sensitive to the imaginary part of $T_{AB}$. Our main results concern the 2-imagitivity; when we speak of imagitivity without specifying the value of $p$, the case $p=2$ will be tacitly understood.

Important properties of imagitivity are captured by \textbf{Theorems} \ref{th1} and \ref{th2} below. 
These apply to any finite-dimensional quantum system undergoing unitary evolution $U(t)$. (Here $U(t)$ is an arbitrary unitary operator, not necessarily governed by a local or time-independent Hamiltonian. Also, note that although RQFT provides motivation for the definition of $T_{AB}$, our theorems apply to finite-dimensional systems.) 
$A$ and $B$ represent two different subsystems at times $0$ and $t$ respectively.
We will concentrate on the case of bosonic operators $\Oc_A, \Oc_B$, meaning that they commute at equal times $[\Oc_A(0), \Oc_B(0)]=0$.
For bosonic operators we can directly apply the definition from Section \ref{sec:def} because $\Oc_{A/B}$ are supported only on $A/B$, whereas fermionic operators are, strictly speaking, non-local as illustrated by the Jordan--Winger transformation. \AM{Ref. \cite{Diaz:2023npx, Diaz:2025aqe} describes a similar construction for the fermionic case.}
When discussing local systems, $|A|$ will denote the number of lattice sites.

\begin{theorem}
\label{th1}
For any bosonic operators $\Oc_A$, $\Oc_B$ supported on $A$ and $B$ respectively:
\beq
\label{eq:th1}
\frac{| \bra [\Oc_A(0), \Oc_B(t)] \ket|}{||\Oc_A||_2 ||\Oc_B||_2} \le || T_{AB} - T_{AB}^\dagger ||_2.
\eeq
\end{theorem}
This is a particular case of eq.(\ref{eq:bound}).
\begin{theorem}
\label{th2}
Given $T_{AB}$, there exist two bosonic operators $\Oc_A, \Oc_B$ such that
\beq
\label{eq:th2}
\frac{1}{\dim \Hc_A} ||T_{AB}-T_{AB}^\dagger||_2  \le \frac{| \bra [\Oc_A(0), \Oc_B(t)] \ket|}{||\Oc_A||_2 ||\Oc_B||_2}.
\eeq
\end{theorem}
These theorems demonstrate that the region $A$ can influence $B$ if and only if the imagitivity $||T_{AB}-T_{AB}^\dagger||_2$ is nonzero.
To make the bound in \textbf{Theorem} \ref{th2} slightly tighter, one can substitute $\dim \Hc_A$ by 
$\min(\dim \Hc_A, \dim \Hc_B)$.
We prove \textbf{Theorem} \ref{th2} in App.~\ref{app:bounds-commutators}. We note that the operators $\Oc_{A,B}$ in this theorem depend on $T_{AB}$. If we change $T_{AB}$ by changing the amount of time-evolution, the corresponding $\Oc_{A,B}$ may change, too.

Our bounds involve 2-imagitivity. A convenient property of this quantity is that its square can be written as
\beq
||T_{AB}-T_{AB}^\dagger||^2_2 =
2\Tr(T_{AB} T_{AB}^\dagger)-2 \Re \Tr(T_{AB}^2) .
\eeq
Hence it can be measured directly by acting on two copies of the system. We present the measurement protocols in Sec.~\ref{sec:measure}.
We obtain a looser lower bound by noticing that $\Tr( T_{AB} T_{AB}^\dagger) \ge |\Tr T_{AB}^2|$. Hence nonzero $\Im \Tr T_{AB}^2$  or negative $\Re \Tr T_{AB}^2$ implies that regions $AB$ are causally connected:
\begin{corollary}
\label{col:im_bound}
Given $T_{AB}$, there exist two bosonic operators $\Oc_A, \Oc_B$ such that
\beq
\label{eq:im_bound}
\frac{2}{\dim \Hc_A} \l(
|\Tr (T_{AB}^2)| - \Re \Tr( T_{AB}^2)
\r) \le \frac{| \bra [\Oc_A(0), \Oc_B(t)] \ket|}{||\Oc_A||_2 ||\Oc_B||_2}.
\eeq
\end{corollary}
Thus $\Im \Tr T_{AB}^2 \neq 0$ or $\Re \Tr T_{AB}^2 < 0$ implies a causal connection between $A$ and $B$.

The converse is not true: $\Im \Tr T_{AB}^2=0$ does not imply $||T_{AB}-T_{AB}^\dagger||_2=0$; it is easy to construct counter-examples even for a system of two qubits. And an example in which $\Re \Tr T_{AB}^2 > 0$ but there is a causal connection is realized for free fermions (Figure \ref{fig:ttbar}).

For local Hamiltonian systems, commutators of local operators obey Lieb--Robinson bounds \cite{Lieb:1972wy, Hastings:2005pr, Hastings:2010loc, Roberts:2016wdl,Barch:2024kwt}, which suggests that $||T_{AB} - T_{AB}^\dag||_2$ should obey a similar bound. The precise formulation of the bound depends the locality of the underlying Hamiltonian and the lattice connectivity \cite{Hastings:2005pr,Nachtergaele_2006}. Hence we will invoke it as an assumption:
\begin{corollary}
Suppose that for any bosonic operators $\Oc_A, \Oc_B$ supported on $A,B$:
\beq
\frac{|| [\Oc_A(0), \Oc_B(t)] ||_2}{||\Oc_A||_2 ||\Oc_B||_2} \le 
C  |A| |B| 
e^{-\mu d(A,B)} \l( e^{v|t|} -1 \r),
\label{eq:LR_bound}
\eeq
where constants $C, v, \mu$ only depend on the underlying lattice and the Hamiltonian, and $d(A,B)$ is the distance between subsets $A,B$. 
Then applying H\"older's inequality to the right hand side of eq.(\ref{eq:th2}) we immediately have
\beq
\begin{split}
||T_{AB} - T_{AB}^\dag||_2 \le 
\dim \Hc_A \frac{|\Tr \l( \rho [\Oc_A(0), \Oc_B(t)] \r)|}{||\Oc_A||_2 ||\Oc_B||_2} \le \\
\le \dim \Hc_A \frac{||\rho||_2 ||[\Oc_A(0), \Oc_B(t)]||_2}{||\Oc_A||_2 ||\Oc_B||_2}
\le \\
\le \dim \Hc_A C |A| |B| 
e^{-\mu d(A,B)} \l( e^{v|t|} -1 \r).
\end{split}
\label{eq:T_LR_bound}
\eeq
\label{col1}
\end{corollary}
Inequality (\ref{eq:LR_bound}) can be proven under very mild conditions on the locality of the Hamiltonian \cite{Hastings:2005pr,Nachtergaele_2006}. 
We will illustrate the bound (\ref{eq:T_LR_bound}) numerically for an Ising chain in Sec.~\ref{sec:numerical-ising} (see Figure \ref{fig:commutators_bound}).

One can also wonder if we can obtain a tight bound on the commutator.
Using $T_{AB}$, we can define a hermitian operator $M=i(T_{AB} - T_{AB}^\dag)$.
Now we perform a realignment \cite{Rudolph_2005,chen2003matrixrealignmentmethodrecognizing}, obtaining from $M$ a map $M_T : \Hc_A \otimes \Hc_A^* \ra \Hc_B \otimes \Hc_B^*$. 
with matrix elements defined by
\beq 
\label{eq:MT}
(M_T)_{aa'}^{bb'}= M_{ab}^{a'b'}.
\eeq
A similar operation is performed to obtain the so-called computable cross-norm \cite{Yin:2022toc,Milekhin:2022zsy}.
In App.~\ref{app:bounds-commutators}, we prove the following
\begin{theorem}
\label{th3}
For any bosonic operators $\Oc_A, \Oc_B$
\beq 
\frac{|\langle [ \Oc_A(0), \Oc_B(t) ] \rangle|}{||\Oc_A||_2 ||\Oc_B||_2} \le  ||M_T||_\infty.
\label{b1}
\eeq
Moreover the bound is tight: there exist $\Oc_A, \Oc_B$ for which the inequality is saturated.
\end{theorem}

In the rest of the paper we will study $T_{AB}$ for various systems. For brevity, we will drop the subscript $AB$ and just write $T$. 

\section{Numerical evaluation for an Ising chain}
\label{sec:numerical-ising}
We consider a quantum chaotic Ising chain with the Hamiltonian
\beq
\label{eq:ising-ham}
H = J \left( \sum_{i = 1}^n Z_i Z_{i+1} + h \sum_{i=1}^n X_i + B_z \sum_{i=1}^n Z_i \right),
\eeq
and with periodic boundary conditions $(Z_{n+1} = Z_1)$, where $Z_i, X_i$ are the Pauli $Z$ and $X$ operators, respectively, acting on the $i$th qubit.

\begin{figure}[h] 
\includegraphics[width=1.05\linewidth]{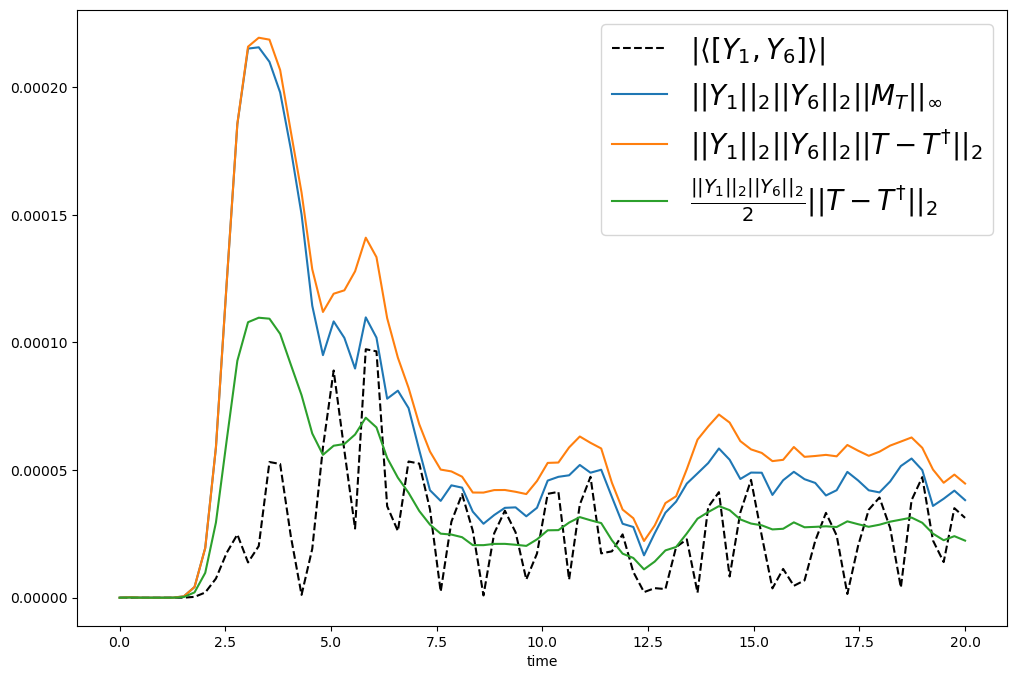}
\caption{Numerical demonstration of inequalities (\ref{eq:th1}), (\ref{lg7}), (\ref{b1}) for the Hamiltonian eq.~\ref{eq:ising-ham} and Pauli Y operators acting on the first and sixth qubits at different times. The plot was obtained for a quantum Ising model with 11 spins and parameters $J = 1, h = -1.05, B_z = 0.5$, and a thermal initial state $\rho_0$ with the temperature $100 J$.}  
\label{fig:commutators_bound} 
\end{figure}

Figure \ref{fig:commutators_bound} demonstrates the inequalities (\ref{eq:th1}) and (\ref{b1}) for Pauli Y operators acting on the first and sixth spins in an 11-spin Ising chain at different times. The dashed line represents the actual expectation value of the commutator. The blue line is the upper bound on the commutator obtained from $M_T$ via (\ref{b1}), and the orange line is the looser upper bound on the commutator obtained from $||T - T^{\dag}||_2$ via (\ref{eq:th1}). The green line is a lower bound on the blue line, as follows from eq.~\ref{lg7} in App.~\ref{app:bounds-commutators}.
Note that both the commutator and $||T - T^{\dag}||_2$ start increasing from zero after a sufficiently long elapsed time, as predicted by Lieb-Robinson bounds. 

We have chosen a relatively large temperature compared to the Hamiltonian coupling strength to test how tight our upper bounds are. At infinite temperature the commutator is identically zero and at large temperature it must be small. As seen in Figure \ref{fig:commutators_bound}, both the commutator and $||T-T^\dagger||$ are of the same order, roughly $10^{-4}$, when the temperature is 100 times larger than the coupling $J$.

\section{Free fermions}
\label{sec:free-fermions}
In this section we consider a one-dimensional free-fermion chain with the Hamiltonian:
\beq
\label{eq:RH}
H = \sum_s \psi(s) \bar{\psi}(s+1)  + h.c.
\eeq
We will concentrate on the case of half-filling and the vacuum state.
Since correlators obey Wick's theorem, all reduced density matrices are Gaussian \cite{Peschel_2003,Sorkin:2012sn,Saravani:2013nwa,Chen:2020ild}. This allows efficient numerical computations of entropies using single-particle representations. 
This is also true for $T$; it is a Gaussian operator because time-separated correlation functions also obey Wick's theorem. This observation is helpful because the microscopic definition of $T$ described in Sec.~\ref{sec:def} does not directly apply to fermionic systems; instead, we extract $T$ from two-point correlation functions.
\begin{figure}
    \centering
    \includegraphics[width=0.9\linewidth]{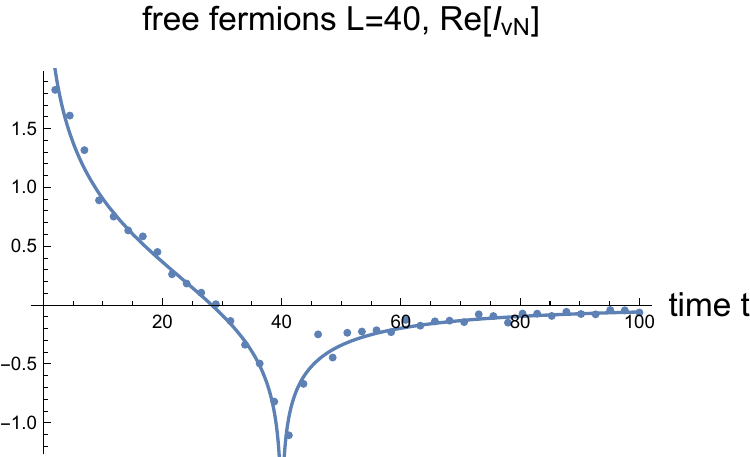}
    \caption{Real part of the mutual information $I_{vN} = S_{vN}(AB) - S_{vN}(A) - S_{vN}(B)$ for free fermions. Here $S_{vN}(A), S_{vN}(B)$ are the conventional von Neumann entropies for intervals $A,B$: $(c/3) \log L/\ep$.
    Dots are the numerical lattice result obtained from taking the principal branch of $\log T$, and the solid line is the CFT prediction $(1/3) \Re \log(1-L^2/t^2)$ obtained from eq.(\ref{eq:two_int_renyi}).
    We used two intervals $[0,L]$ of length $L=40$ but separated in time by $t$.
    }
    \label{fig:reI}
\end{figure}

\begin{figure}
    \centering
    \includegraphics[width=0.9\linewidth]{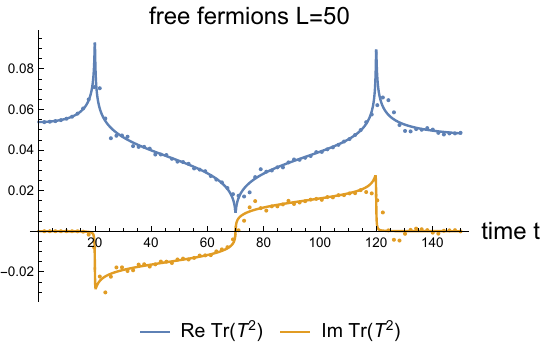}
    \caption{The real and imaginary parts of $\Tr T^2$ for free fermions. Dots are the numerical lattice result, solid lines are CFT predictions; see eq.(\ref{eq:cft_tt}). 
    Subsystems $A$ and $B$ both have length $L=50$, but $A$ is the interval $[0,50]$ and
    subsystem $B$ is the interval $[70,120]$ which is moved forward in time by $t$. }
    \label{fig:ttbar}
\end{figure}

In App.~\ref{app:free-fermion} we explain how to use standard free-fermion techniques to extract $T$. Figures \ref{fig:reI}, \ref{fig:ttbar} show the results for the mutual information and $\Tr T^2$, assuming $A,B$ denote the same spatial interval of length $L$, but separated by time $t$. 
It is well-known that in the infrared this system can be described by the $c=1$ complex free fermion CFT. The first two quantities can be easily obtained from the known answer for the Tsallis entropies \cite{Casini:2005rm}. Using Euclidean complex coordinates $z=s+i\tau$ for the two intervals $[z_1, z_2] \cup [z_3,z_4]$, the answer is
\beq
\begin{split}
&\Tr \rho^n =  \\
&=\l(  \frac{ \ep^4}{ x \bar{x} (z_2-z_1) (\bar{z}_2-\bar{z}_1) (z_4-z_3) (\bar{z}_4-\bar{z}_3)} \r)^{-\frac{n-1}{12n}}, 
\end{split}
\label{eq:two_int_renyi}
\eeq
where $x$ is the conformal cross-ratio
\beq
x =\frac{(z_4-z_1)(z_3-z_2)}{(z_3-z_1)(z_4-z_2)},
\label{eq:CR}
\eeq
and $\ep$ is a short-distance cutoff.
To obtain $\Tr T^n$ we perform the (standard in RQFT \cite{peskin2018introduction}) continuation from $(z,\bar{z})$ to the light-cone coordinates $(u,v)=(s-t,s+t)$, where $s,t$ are conventional real space and time coordinates. By giving time coordinates of the interval $A$ a small positive imaginary part $+ i \varepsilon$, and time coordinates of $B$ a small negative imaginary part $-i\varepsilon$, we enforce that $A$ precedes $B$ in the operator ordering. This resolves the ambiguity coming from the branch-cut of the fractional power in (\ref{eq:two_int_renyi}).
 Figures \ref{fig:reI}, \ref{fig:ttbar} illustrate that the numerical results for the mutual information and for $\Tr T^2$ agree well with predictions obtained by analytic continuation in RQFT. 

We mentioned in section \ref{sec:properties} that $\log T$ can be unambiguously defined if we start from the standard space-like separated $A,B$ can continuously deform to time-like separation. 
Unfortunately, this approach is technically demanding to implement in practice and Figure \ref{fig:reI} was obtained by taking the principal branch of the logarithm instead. The agreement with the CFT computation suggests that at least for free fermions the eigenvalues do not acquire large phases.

\section{$(1+1)$-dimensional conformal field theories}
\label{sec:cft}
In a generic CFT, the entropy of a pair of disjoint intervals is not universal.
However, with two replica copies the answer can be mapped to the torus partition function.
In this section we will perform this mapping to compute $\Tr(T^2)$.

A plane with two slits, where the intervals $A=[z_1,z_2],B=[z_3,z_4]$ are, can be mapped to a cylinder. Gluing two such cylinders together to obtain $\Tr T^2$ or $\Tr \rho_{AB}^2$, we end up with a torus geometry. Equivalently, this computes the four-point function of twist operators. Dedicating the details to App. \ref{app:cft2}, we quote the answer here:
\begin{align}\label{eq:cft_tt}
\Tr T^2 &= Z_2(\tau,\bar{\tau}) 2^{-2c/3} \times \nonumber  \\
&\times \l( \frac{\ep^4}{(z_2-z_1) (\bar{z}_2-\bar{z}_1) (z_4-z_3) (\bar{z}_4-\bar{z}_3) } \r)^{1/8} \times \nonumber  \\
&\times \l( \frac{x}{(1-x)^2} \r)^{-c/24}
\l( \frac{\bar{x}}{(1-\bar{x})^2} \r)^{-c/24},
\end{align}
where $c$ is the central charge, $Z_2(\tau,\bar{\tau})$ is the CFT torus partition function which depends on the modular parameter $\tau$, and the extra factor comes from the conformal anomaly (the Liouville action), because the two-replica geometry maps to a torus with non-flat metric \cite{Lunin:2000yv}. Cross-ratio $x$ and modular parameter $\tau$ are related by:\footnote{We are using Wolfram Mathematica conventions for the theta-functions and $K$ is the complete elliptic integral of the first kind.}
\beq
x= \frac{\theta_2(e^{2 \pi i \tau})^4}{\theta_3(e^{2 \pi i \tau})^4}, \ \tau = \frac{i}{2} \frac{K(1-x)}{K(x)},
\eeq
\beq
\bar{x}= \frac{\theta_2(e^{-2 \pi i \bar{\tau}})^4}{\theta_3(e^{-2 \pi i \bar{\tau}})^4}, \ \bar{\tau} = -\frac{i}{2} \frac{K(1-\bar{x})}{K(\bar{x})}.
\eeq
The cross-ratio $x$ is given by eq.(\ref{eq:CR}) and for $\bar{x}$ one needs to substitute all $z$ by $\bar{z}$. Importantly, when we promote $z,\bar{z}$ to the light-cone coordinates, $\tau$ and $\bar{\tau}$ are no longer complex conjugate of each other. Technically, this breaks the modular invariance and we provide further comments on this in Appendix \ref{app:cft2}.

As an example, for complex Dirac fermion the torus partition function is \cite{francesco2012conformal}:
\beq
Z^{\text{free fermions}}_2(\tau,\bar{\tau}) = \frac{\theta_3(e^{2 \pi i \tau}) \theta_3(e^{-2 \pi i \bar{\tau}})}{\eta(2 \tau) \eta(-2 \bar{\tau})}.
\label{eq:ff_z}
\eeq
Using the theta-function identities from Appendix \ref{app:free-cft}, one can show that eq.(\ref{eq:cft_tt}) reduces to eq.(\ref{eq:two_int_renyi}) for $n=2$.

\subsection{A relation to timelike pseudoentropy}
\label{subsec:pseudoentropy}
References \cite{Mollabashi:2020yie,Nakata:2020luh,Doi:2023zaf,Doi:2022iyj} introduced the notion of timelike pseudoentropy for RQFTs: compute the entanglement entropy for a spacelike interval and analytically continue it to make the interval timelike. Refs. \cite{Carignano:2024jxb,Bou-Comas:2024pxf} extended this proposal to local tensor networks and discussed possible measurement protocols.
However, it is not clear whether timelike pseudoentropy defined this way can be generalized beyond these two cases.
Notice that it is distinct from the case studied in this paper, since we take two intervals which are both spacelike, and then allow the separation between the two intervals to be timelike.
The entanglement in time we defined is applicable to any system and in certain cases it coincides with the timelike pseudoentropy. In this sense our construction provides a microscopic and measurable definition for the timelike pseudoentropy.

\begin{figure}
    \centering
    \includegraphics[scale=1.2]{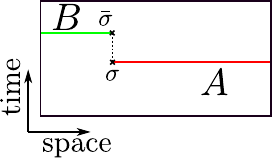}
    \caption{Two timelike separated twist operators.}
    \label{fig:two_twist}
\end{figure}

The key point is that both pseudoentropy and entanglement in time can be formulated in terms of correlation functions of timelike separated twist operators.
In the particular case of a $(1{+}1)$-dimensional CFT in infinite volume a correlation function of two twist operators, Figure \ref{fig:two_twist}, can be interpreted as either computing the pseudo-entanglement entropy for a single timelike interval (black, dashed) \textit{or} entanglement in time (that is, $-\Tr T \log T$) for two spacelike intervals $A,B$ separated in the time direction (green and red). 

\section{Holography}
Other situations where we can compute the entanglement in time arise in the holographic systems --- quantum systems which have a dual gravity description in higher dimensions. Then von Neumann entropy can be computed by finding the extremal surface in the bulk anchored at the corresponding boundary region \cite{Ryu:2006bv,Hubeny:2007xt}.
As emphasized in Sec.~\ref{subsec:pseudoentropy}, entanglement in time can be related to the correlation function of time-separated twist operators, so in some cases the holographic computation is identical to the computation of timelike pseudoentropy.
Additionally, since we are computing an RQFT correlation function, our answers must be consistent with the operator product expansion (OPE) when various twist operators are close to each other in Lorentzian distance. This is straightforward when a twist and an anti-twist operator become close, but it is more subtle for two twist operators because of how topological defect lines can attach to them. We will explain this in detail in Appendix \ref{app:ope}.

Now the extremal surface in the bulk must be anchored at the time-separated twist operators, indicating that at least a portion of the surface is timelike. However, there is a subtlety --- such \textit{real} timelike extremal surfaces might not exist. 
The choice of \textit{complex} extremal surface is subtle; we refer to \cite{Heller:2024whi} for a recent discussion. We postpone a more complete analysis for future work.
For now we discuss a simple case of two intervals and provide some intuition about the complex extremal surfaces relevant to the computation of $T$, which differ from those arising in the computation of timelike pseudoentropy.

Specifically, for ergodic systems (including the holographic systems \cite{Hayden:2007cs,Sekino:2008he,Shenker:2013pqa,Bizon:2011gg}) for large time separation we expect all two-point functions to factorize, hence we conjecture that $T$ factorizes as well:
\beq
\label{eq:factorization}
T \approx \rho_A \otimes \rho_B, \ \Delta t \ra +\infty.
\eeq
\begin{figure}
    \centering
\includegraphics{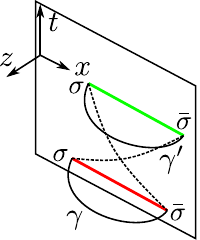}
    \caption{Poincar\'e patch $AdS_3$ with two intervals at the boundary. Geodesic $\gamma$ (solid) is a real extremal surface, whereas $\gamma'$ (dashed) is imaginary.}
    \label{fig:rt_time}
\end{figure}

For simplicity we will focus on a $(1{+}1)$-dimensional holographic CFT in a vacuum state in infinite volume. 
The corresponding bulk dual is a Poincar\'e patch of three-dimensional anti-de Sitter (AdS) geometry with the metric 
\beq
ds^2 = \frac{-dt^2 + dz^2 + dx^2}{z^2}.
\eeq
The boundary regions are two intervals of length $L$ on top of each other separated by time $\Delta t$ as shown in Figure \ref{fig:rt_time}.
There are two extremal surfaces (in $2+1$ dimensions they are geodesics): purely real spacelike $\gamma$ consisting of two geodesics of the form:
\beq
\sqrt{z^2 + x^2} =  \frac{L}{2}.
\eeq

The corresponding entanglement entropy is $\frac{1}{4G_N}$ times twice the length of the geodesic: $\frac{2c}{3} \log(L/\ep)$ --- this is simply the sum of two entanglement entropies of $A$ and $B$. From the factorization property (\ref{eq:factorization}) we expect this contribution to dominate at late times (large $\Delta t$). Also, it correctly reproduces the OPE limit when $A$ and $B$ are very small.

One can easily obtain the \textit{imaginary} surface $\gamma'$ connecting the twist $\sigma$ and anti-twist $\bar{\sigma}$ between $A$ and $B$ --- see Figure \ref{fig:rt_time}.
The real part of its doubled-length is $\frac{2c}{3} \log{(\sqrt{L^2-(\Delta t)^2}/\ep)}$. 
How do we determine whether $\gamma$ or $\gamma'$ dominate?
Naively, we should look for the minimal real part.
This prescription selects
$\gamma'$ ($\frac{2c}{3} \log{(\sqrt{L^2-(\Delta t)^2}/\ep)}$) for $\Delta t< \sqrt{2} L$ and $\gamma$ ($\frac{2c}{3} \log(L/\ep)$) for $\Delta t > \sqrt{2} L$, in accord with the conjecture (\ref{eq:factorization}). 

Potentially, there is another imaginary surface that connects the twist operator $\sigma$ of $A$ to the twist operator $\sigma$ of $B$ (and similarly for $\bar{\sigma}$). However, for spacelike separation such a surface does not satisfy the holographic homology condition, so we expect that this surface is not relevant.

The minimal real part prescription seems to work in this particular case; moreover it correctly reproduces the OPE limit when the endpoints of $A$ and $B$ are close to null-separation.
However, in more complicated examples at finite temperature it might produce non-physical results \cite{Heller:2024whi}. It would be interesting to revisit such examples from the microscopic perspective of timelike entanglement entropy.

\section{Measurement protocols}
\label{sec:measure}
Here we discuss protocols for measuring $\Tr T^2$ and $\Tr T T^\dagger$. From these quantities we can obtain the 2-imagitivity $
2\Tr T T^\dagger -2 \Re \Tr T^2 $. These measurements are performed acting on two identical copies of the quantum system. We used a variant of a SWAP test, but it would also be interesting to explore other approaches, for example ref. \cite{Bou-Comas:2024pxf} proposed a quench protocol for measuring temporal entanglement and \AM{ref. \cite{Diaz:2021snw} considered circuits with controlled unitary evolution.}

\begin{figure}
    \centering
    \includegraphics{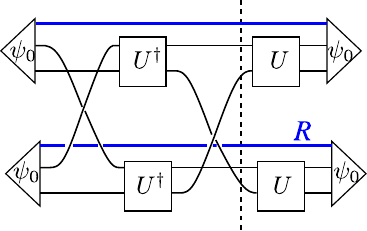}
    \caption{Tensor diagram representing $\Tr T^2$, separated by the dashed line into $\langle \chi|$ and  $| \phi \ket$. The thick blue line indicates the reference system $R$ purifying the density matrix $\rho$ as in eq.(\ref{eq:rhoR}).}
    \label{fig:TrTT}
\end{figure}

The state $\rho$ may be mixed in general, but we will consider its purification $|\psi\rangle_0$ by a reference system $R$; that is, 
\beq
\rho = \Tr_R \left(|\psi_0\rangle\langle \psi_0|\right).
\label{eq:rhoR}
\eeq
Comparing to eq.~(\ref{eq:TAB}) one sees that $\Tr T^2 $ can be represented by the tensor diagram shown in Figure \ref{fig:TrTT}. This tensor contraction can be interpreted as the overlap of two pure states:
\beq
\Tr T^2 = \bra \chi | \phi \ket
\eeq
where
\beq
| \phi \ket = U^{\otimes 2} |\psi_0 \ket^{\otimes 2},
\eeq
and
\beq
| \chi \ket = \text{SWAP}_{B_1B_2} U^{\otimes 2} \text{SWAP}_{A_1A_2} |\psi_0 \ket^{\otimes 2}.
\eeq
Here $\text{SWAP}_{A_1A_2}$ swaps the $A$ subsystems of the two copies of the state and $\text{SWAP}_{B_1B_2}$ swaps the $B$ subsystems. Because $\Tr T^2$ is the inner product of two normalized states, this representation proves that $|\Tr T^2|\le 1$, as announced earlier. Similar arguments lead to $|\Tr T^n| \le 1$ for $n > 2$ (Property 4 stated in Sec.~\ref{sec:properties}). 

The overlap of two pure states can be computed using the standard SWAP test with a single ancilla qubit. However using this method we require both $|\phi\rangle$ and $|\chi\rangle$ as input states, and therefore all together four copies of the state $|\psi_0\rangle$. Fortunately, for evaluating this particular overlap, only two copies of $|\psi_0\rangle$ are actually needed. 

\begin{figure}
    \centering
\includegraphics{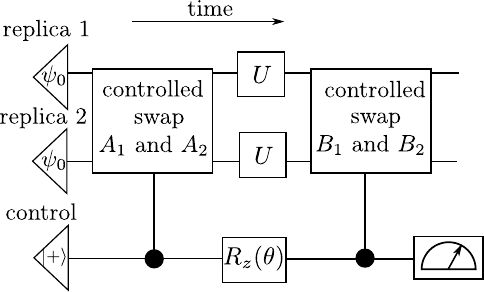}
    \caption{Measurement protocol for $\Tr T^2$.}
    \label{fig:curcuit_TT}
\end{figure}

To measure the overlap $ \langle \chi | \phi \rangle$, we construct a circuit consisting of two copies of the initial state $|\psi_0\ket$ and an ancillary qubit in the state $|+\ket$. Thus, the initial state of the system can be expressed as
\beq
|+\ket \otimes |\psi_0\ket^{\otimes 2} = \frac{1}{\sqrt{2}} \left( |0 \ket \otimes |\psi_0\ket^{\otimes 2} + |1\ket \otimes |\psi_0\ket^{\otimes 2} \right).
\eeq
We first apply a controlled-SWAP gate, which swaps the subsystems $A$ between the two copies condition on the state of the ancillary qubit,
\beq
\text{CSWAP}_{A_1A_2} = |0\ket \otimes \id + |1\ket \otimes \text{SWAP}_{A_1A_2}.
\eeq
Then, we apply a rotation $R(\theta) = e^{i\frac{\theta}{2}Z}$ by angle $\theta$ about the $z$-axis to the ancillary qubit, evolve the two copies of the system with a unitary $U$, and finally apply CSWAP$_{B_1B_2}$ --- a controlled-SWAP gate on the $B$ subsystems within the two copies --- as shown in Figure \ref{fig:curcuit_TT}. This results in the final state
\beq
\begin{split}
&|\Psi\rangle =\frac{1}{\sqrt{2}} ( e^{i \frac{\theta}{2}}|0 \ket \otimes U^{\otimes 2} |\psi_0 \ket^{\otimes 2} \\
& + e^{-i \frac{\theta}{2}}|1\ket \otimes \text{SWAP}_{B_1B_2} U^{\otimes 2} \text{SWAP}_{A_1A_2} |\psi_0 \ket^{\otimes 2} )  \\
&= \frac{1}{2} \left[|+\ket (e^{i \frac{\theta}{2}}|\phi\ket + e^{-i \frac{\theta}{2}}|\chi\ket) + |-\ket (e^{i \frac{\theta}{2}}|\phi\ket - e^{-i \frac{\theta}{2}}|\chi\ket) \right].
\end{split}
\eeq
Hence, when we measure the ancillary qubit in the $|+\rangle$, $|-\rangle$ basis, the measurement outcomes occur with probabilities
\begin{align}
\mathbb{P}(|+\ket) &= \frac{1}{4}\left(2 + e^{i \theta} \langle \chi|\phi\rangle +e^{-i \theta}\langle \phi|\chi\rangle\right),\\
\mathbb{P}(|-\ket) &= \frac{1}{4}\left(2 - e^{i \theta} \langle \chi|\phi\rangle -e^{-i \theta}\langle \phi|\chi\rangle\right),
\end{align}
and therefore
\begin{align}
 \mathbb{P}(|+\ket) -\mathbb{P}(|-\ket)=\frac{1}{2}\left(e^{i\theta}\left(\Tr T^2\right) + e^{-i\theta} \left(\Tr T^2\right)^*\right) .
\end{align}
Thus we obtain the real part of $\Tr T^2$ for $\theta=0$ and the imaginary part for $\theta = - \pi/2$. By performing sufficiently many trials to estimate the outcome probabilities accurately for both $\theta = 0$ and  $\theta=-\pi/2$, we can determine the real and imaginary parts of $\Tr T^2$.

\begin{figure}
    \centering
\includegraphics{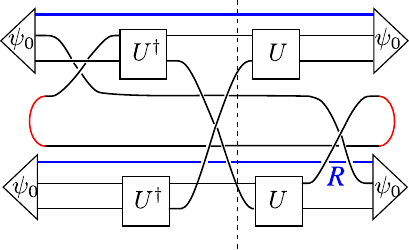}
    \caption{Tensor diagram representing $\Tr(T T^\dagger)$, separated by the dashed line into $\langle \chi'|$ and $|\phi'\rangle$.  The red arcs indicate auxiliary EPR pairs, and the thick blue line indicates the reference system $R$.}
    \label{fig:TrTTd}
\end{figure}

\begin{figure}
    \centering
    \includegraphics{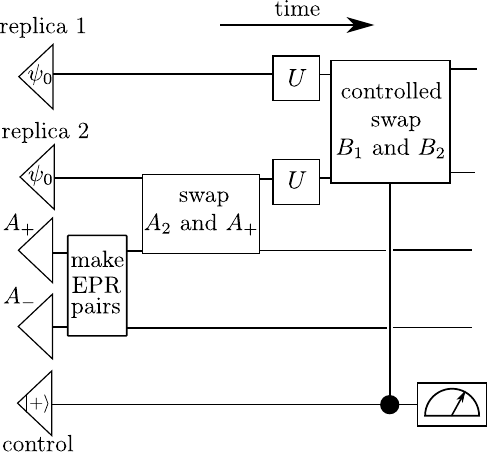}
    \caption{Measurement protocol for $\Tr T T^\dagger$.}
    \label{fig:circuit_TTd}
\end{figure}

The quantity $\Tr T T^{\dag}$ also can be expressed as a tensor contraction, a more complicated one depicted in Figure \ref{fig:TrTTd}. It, too, can also be interpreted as a state overlap, but now we need to introduce two additional subsystems, $A_+$ and $A_-$, which are prepared in a maximally entangled state and measured in a maximally entangled basis, as shown in Figure \ref{fig:TrTTd} (red chords). Here $A_+$ and $A_-$ have the same dimension as $A$. Because the maximally entangled state
\begin{equation}
|\Phi\rangle_{A_+A_-}=\frac{1}{\sqrt{\dim \Hc_A}}\sum_a |a\rangle\otimes|a\rangle
\end{equation}
is normalized, a factor of $\frac{1}{\dim \Hc_A}$ enters the computation when we write the tensor contraction as a state overlap. Therefore we have
\beq
\frac{1}{\dim \Hc_A}\Tr T T^{\dag} =\langle \chi' | \phi' \rangle,
\eeq
where
\beq
|\phi'\rangle = \left(U\otimes I\otimes U\right)\text{SWAP}_{A_+A_2}\left(|\psi_0\rangle \otimes|\Phi\rangle\otimes |\psi_0\rangle\right),
\eeq
and 
\beq
|\chi'\rangle = \text{SWAP}_{B_1B_2}|\phi'\rangle.
\eeq
The quantum circuit shown in Figure \ref{fig:circuit_TTd} prepares the state $|\phi'\rangle$ if the control qubit is $|0\rangle$ and the state $|\chi'\rangle$ if the ancillary control qubit is $|1\rangle$. Therefore, right before the ancillary qubit is measured we have the state 
\begin{align}
|\Psi\rangle =\frac{1}{\sqrt{2}}\left(|0\rangle \otimes |\phi'\rangle + |1\rangle\otimes |\chi' \ket \right).
\end{align}
When the ancillary qubit is measured in the $|+\rangle$, $|-\rangle$ basis, the outcome $|+\rangle$ occurs with probability
\beq
\begin{split}
 \mathbb{P}(|+\ket) &= \bra +| \Psi \ket\bra \Psi| +\ket  = 
\frac{1}{4} \left(\bra\phi'| + \bra\chi'|\right) \left(|\phi'\ket + |\chi'\ket\right)  \\
&=\frac{1}{2} \left( 1 + \frac{1}{\dim \Hc_A} \Tr  TT^{\dag} \right) .
\end{split}
\label{eq:TTd_prob}
\eeq

To demonstrate this in practice, we implemented these measurement protocols for a single qubit with zero Hamiltonian, with initial density matrix $\rho = |+\ket \bra+|$.  Here $A$ and $B$ refer to the same qubit, but at different ``times.'' 
Explicit calculation of $T_{AB}$ for this system yields $\Tr T^2  = 1$ and $\Tr TT^\dag  = 2$. Table \ref{table:qiskit} shows results obtained with the qiskit \texttt{Aer} simulator (with \texttt{ibm\_sherbrooke} backend), the \texttt{ibm\_sherbrooke} device without applying any error mitigation, and the \texttt{ibm\_sherbrooke} device with with error mitigation strategies outlined below \cite{qiskit2024}  \cite{ibmquantum2021}.
\begin{table}
\centering
\begin{tabular}{|c|c|c|c|c|}
\hline
 & \textbf{Theory} & \textbf{Simulation} & \textbf{QPU} & \textbf{\vtop{\hbox{\strut QPU w/ error}\hbox{\strut mitigation}}}\\ \hline
$\Re(\Tr T^2) $ & 1 & 0.882(2) & 0.630(5)& 0.96(9) \\ \hline
$\Im(\Tr T^2)$ & 0 & 0.002(2) & 0.699(4) & $0 \pm 0.04$ \\ \hline
$\Tr T T^{\dag}$ & 2 & 1.79(5)& 1.50(1) & 1.82(6)\\
\hline
\end{tabular}
\caption{$\Tr(T^2)$ and $\Tr(TT^\dag)$ computed using the proposed measurement protocols. The ``Simulation'' column shows results obtained with the qiskit \texttt{Aer} simulator (with \texttt{ibm\_sherbrooke} backend). 'QPU' column corresponds to results obtained with the \texttt{ibm\_sherbrooke} device (Eagle r3 processor) without applying any error mitigation, and the last column has results from this device with ZNE (with linear extrapolation), TREX and dynamical decoupling. All values were obtained using 40,000 shots and errors are statistical only. Simulations on other Eagle r3 processors yield results within the expected systematic error reported in the main text. }
\label{table:qiskit}
\end{table}
The imagitivity $||T-T^{\dag}||_2^2 = 2\Tr(TT^\dag) - 2\Re \Tr(T^2)$ was found to be $1.7(2)$ using error mitigation techniques, compared to the theoretically expected value $2$.
To compute the expectation values, we used the \texttt{Estimator} primitive with twirled readout error extinction (TREX), zero noise extrapolation (ZNE), gate twirling, and dynamical decoupling \cite{qiskit2024}. 
The relatively large errors were expected due to noise in the circuits, primarily arising from echoed cross-resonance (ECR) entangling two-qubit gates and single-qubit SX gates. For the \texttt{ibm\_sherbrooke} device, the median error rates are \texttt{7.339e-3} for ECR gates and \texttt{2.493e-4} for SX gates. With $44$ SX gates and $21$ ECR gates in the circuits for $\Tr T^2$, we anticipated (absolute) systematic error without error mitigation of at least $0.165$. For the circuit measuring $\Tr TT^{\dag}$, which involved $37$ SX gates and $11$ ECR gates, the expected error was at least  $2 \times 0.090 = 0.180$ (the factor of 2 comes from $\dim \Hc_A=2$ in equation (\ref{eq:TTd_prob})).

\section{Matrix-product states representations and temporal entanglement}
Matrix product states (MPS) and matrix product operators (MPO) are ubiquitous for approximating ground states and thermal states of gapped local Hamiltonians, particularly in one spatial dimension. Given that $T$ is an analogue of a density matrix, it is natural to ask whether it can be represented as an MPO. We postpone answering this question in full generality for future work and instead provide some intuition and numerical data.

\begin{figure}
    \includegraphics[width=1.0\linewidth]{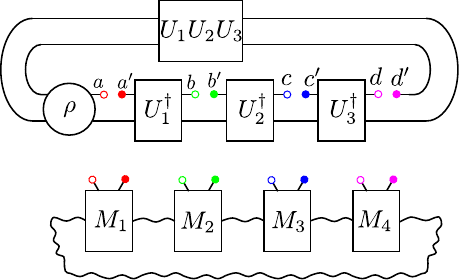}
    \caption{Upper: the generalization of $T$ to include four intervals $ABCD$. Subsystems $A,B,C$ and $D$ are inserted at different times and do not have to be related to each other. Lower: the MPO tensor diagram aiming to approximate $T$. Wiggly lines represent a bond space of dimension $\chi$.}
    \label{fig:TABCD}
\end{figure}

First of all, in the MPO case it is standard to take a large system and cut it into two halves to examine the singular value decomposition of the wave-function/density matrix. 
For $T_{AB}$ the effective size of the generalized density matrix is determined by the spatial size of $A$ and $B$. 
Rather than increasing the size of these subsystems, which would inevitably increase the amount of information related to spacelike correlations, we instead consider increasing the number of time-separated intervals while keeping each interval small (e.g. only a single qubit) to ensure that the singular values characterize time-separated correlations.

It is straightforward to generalize $T$ to include several time-separated intervals. Figure \ref{fig:TABCD} (upper) illustrates the case of four intervals. It is always possible to represent $T$ as an MPO --- see Figure \ref{fig:TABCD} (lower) --- if the bond dimension $\chi$ is sufficiently large. A question of interest is: how does $\chi$ scale for an MPO that accurately approximates $T$?

Typically, MPO representations allow only exponentially decaying correlators, assuming that all the auxiliary tensors ($M_i$ in Figure \ref{fig:TABCD}) are the same and the bond dimension is fixed. Potentially this introduces an obstacle for approximating $T$, since for Hamiltonian systems the late-time dynamics is often governed by hydrodynamics with power-law decaying correlators, suggesting that $\chi$ has to grow with time as well.

To understand what $\chi$ should be, the most straightforward approach is to cut the system into two halves and examine the corresponding singular values. For example, for four intervals (Figure \ref{fig:TABCD}) this means treating the ``past'' subsystems $AB$ as input matrix legs and ``future'' subsystems $CD$ as output legs. An exact MPO representation would imply that only $2 \chi$ of the singular values are nonzero.  

\begin{figure}[htbp]
\centering
\begin{minipage}{\linewidth}
    \centering
   \includegraphics[scale=0.33]{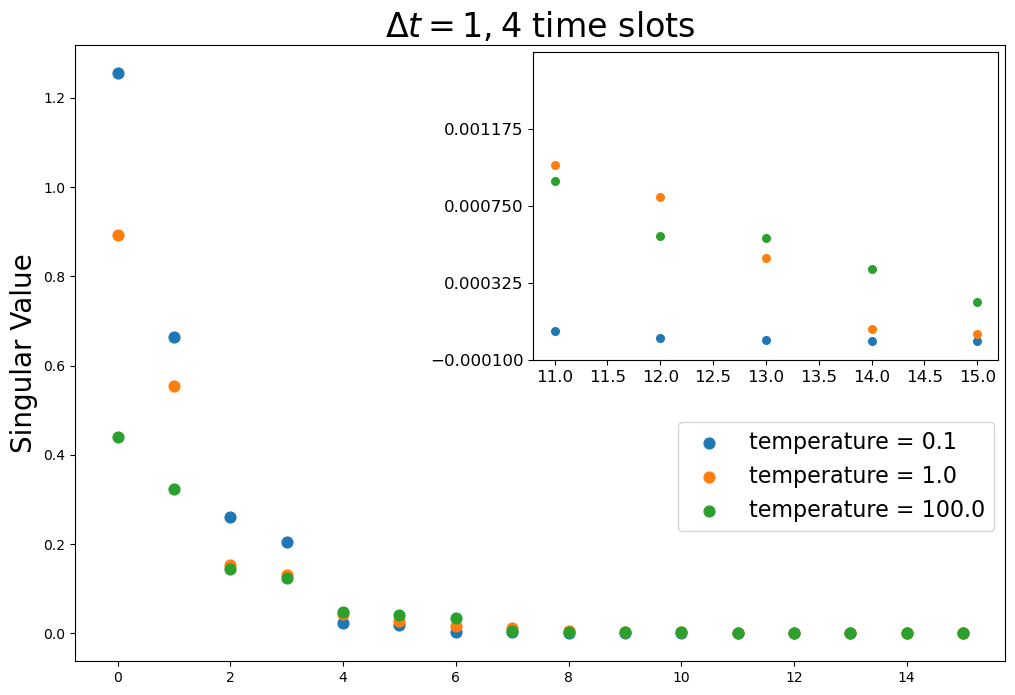} 
\end{minipage}
\begin{minipage}{\linewidth}
\centering
   \includegraphics[scale=0.33]{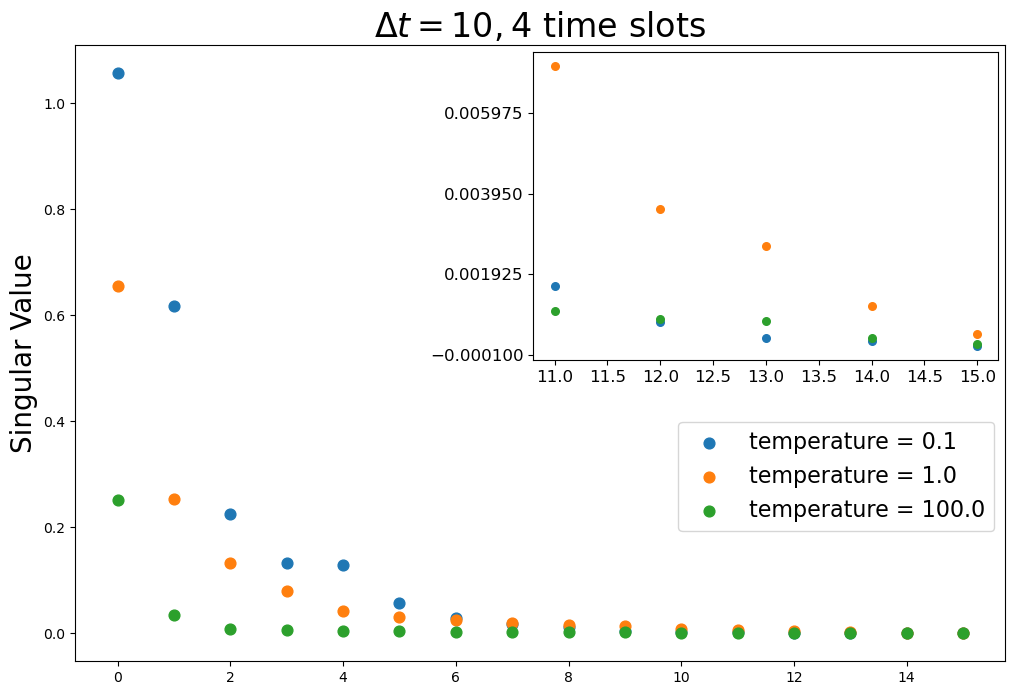} 
\end{minipage}
\begin{minipage}{\linewidth}
\centering
   \includegraphics[scale=0.33]{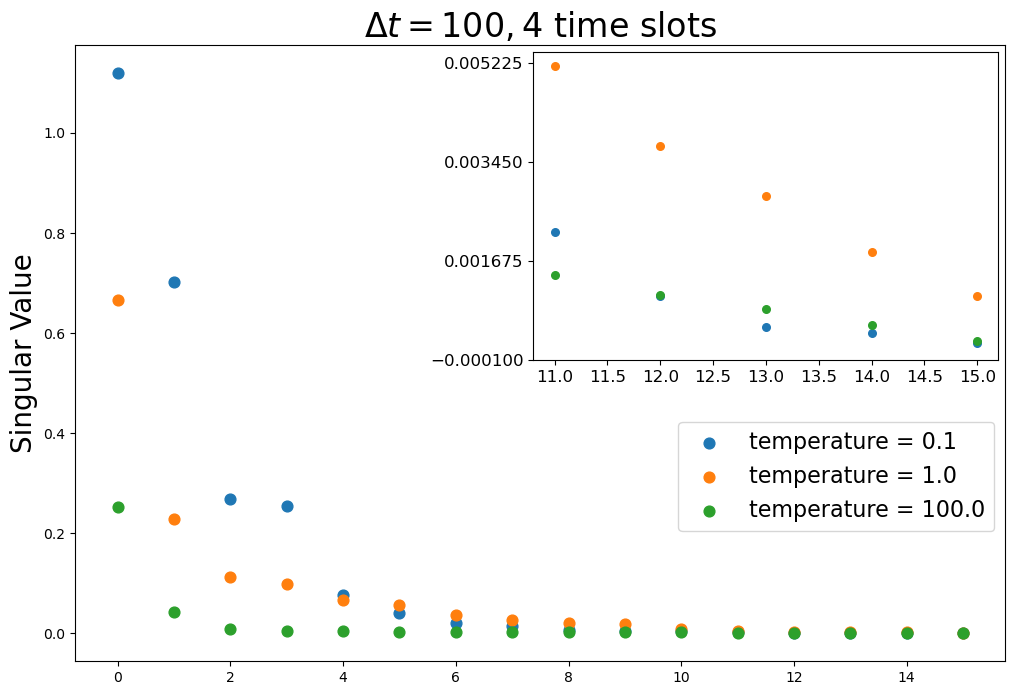} 
\end{minipage}
    
    \caption{Singular values of $T$ for the Ising chain Hamiltonian eq.~\ref{eq:ising-ham} probed at four time slots separated by intervals $\Delta t = 1, 10, 100$. The plot was obtained for a 10-spin chain with parameters $J = 1,h = -1.05,B_z = 0.5$, and a thermal initial state $\rho_0$. }
    \label{fig:MPS}
\end{figure}

In Figure \ref{fig:MPS}, we plot the singular values of a matrix representation of $T$ 
for an Ising chain with 10 spins at different temperatures. This model is known for exhibiting ballistic spread of entanglement and diffusive spread of energy \cite{Kim:2013etb}. We probe the system at four different times, with intervals $\Delta t = 1,10,100$ between them. Each subsystem includes just one spin. Surprisingly, the singular values follow an exponential decay, which suggests that $T$ may be represented as an MPO. It would be interesting to reconcile this observation with power-law decaying correlators.
Recently the possibility of an MPO representation was explored for the Feynman--Vernon influence functional \cite{sonner2021influence,Lerose:2021svg,Thoenniss:2024nlw,Luchnikov:2024qvl} using the notion of temporal entanglement. It would be very interesting to find a connection between that work and our approach, since temporal entanglement captures the complexity of classical simulation of quantum systems \cite{Ba_uls_2009,M_ller_Hermes_2012,Hastings:2014qqa}. 

\section{Conclusion}
In this paper we introduced and studied a generalization of the density matrix that encodes correlations of timelike separated subsystems. Our construction was guided by RQFTs, for which entropies such as $\Tr T^n$ can be obtained via the analytic continuation from spacelike separated regions. However, we provided a general definition of $T$, applicable for \textit{any} quantum system. We have proven several basic properties of $T$ and have shown how it can be computed and measured.

As an illustration of the utility of $T$, we introduced a novel measure, \textit{entanglement 2-imagitivity $||T-T^\dagger||_2$} and showed that it obeys a Lieb--Robinson bound (\textbf{Corollary} \ref{col1}) and provides useful upper bounds (\textbf{Theorem} \ref{th1}) and lower bounds (\textbf{Theorem} \ref{th2}) on commutators of operators. In other words, nonzero imagitivity is necessary and sufficient to signal from $A$ to $B$.
Additionally, we demonstrated that non-zero $\Im \Tr T^2$ or negative $\Re \Tr T^2$ is sufficient for causal contact (\textbf{Corollary} \ref{col:im_bound}).
An appealing feature of these bounds is that they are applicable for arbitrary operators on $A$ and $B$ that are not necessarily local.

Given the numerous applications of the conventional entanglement entropy, we hope that the entanglement entropy in time proposed in this paper will also provide many results and insights.
Since $T$ can capture arbitrarily complicated correlation functions and is amenable to analytic computations, it would be interesting to investigate how its behavior captures integrable versus chaotic dynamics and probes the emergence of hydrodynamics at late times. A specific challenge is finding a timelike generalization of the Bekenstein--Casini bound \cite{bekenstein1981universal,Casini:2008cr}, which restricts the amount of entropy in a given region of space and can be easily derived from the positivity of relative entropy. \AMnew{We refer to \cite{Kull:2018btc} for a work in this direction.} 
Perhaps a similar result pertaining to entanglement in time can be related to the covariant entropy bound \cite{Bousso:1999xy} \AMnew{and extremal time-like surfaces in de Sitter space \cite{Narayan:2015vda,Narayan:2017xca,Doi:2022iyj,Narayan:2022afv,Shaghoulian:2022fop,Narayan:2023zen,Kawamoto:2023nki} and traversable wormholes \cite{Kawamoto:2025oko}  }. 
Likewise, strong subadditivity of von Neumann entropy can be used to prove the $c-$theorem \cite{Casini:2006es}, and it is tantalizing to explore generalizations of this as well. The relation between timelike pseudoentropy and the renormalization group, recently discussed in \cite{Grieninger:2023knz}, may be helpful in this regard. Furthermore given that $\Tr T^n$ in holographic theories is sensitive to bulk (complex) timelike surfaces, one wonders whether it allows us to gain information about the black hole interior. We refer to \cite{Milekhin:2024mce} for a preliminary discussion of this possibility.

While we have provided protocols for measuring $\Tr T^2$ and $\Tr(T T^\dagger)$, we have not fully discussed the operational and information-theoretic properties of $T$. For example, how well does $T$ distinguish the state of the system at time zero and from its evolved state at time $t$?

We note that $T$ as we have defined it does not readily capture measurement-induced dynamics. For that one also needs to account for possible insertions into the backward part of the evolution (the ``minus'' part of the Schwinger--Keldysh contour); properly including measurements will require adding more ``slots'' to $T$.
By design, our definition of $T$ captures Wightman correlation functions as in eq.~(\ref{eq:wightman}). The potential connections relating entanglement in time and imagitivity to out-of-time-order correlators are yet to be explored.

\begin{acknowledgments}
We thank D.~Abanin, A.~Gorsky, A.~Kitaev, M.~Kreshuk, J.~Maldacena, K.~Narayan, S.~Pal, D.~Simmons-Duffin, S.~Solodukhin, J.~Sorce, F.~Surace, S.~Valgushev, I.~Vilkoviskiy, Z.~Wei, Z.~Yang and especially R.~Farrell, S.~Murciano and A.~Serantes for comments and discussions and L.~Tagliacozzo for the comments on the manuscript. 

AM acknowledges funding provided by the Simons Foundation (Grant 376205),  the DOE QuantISED program (DE-SC0018407), and the Air Force Office of Scientific Research (FA9550-19-1-0360). 
ZA acknowledges funding from Robert L. Blinkenberg Summer Undergraduate Research Fellowship. 
JP acknowledges funding provided by the Institute for Quantum Information and Matter, an NSF Physics Frontiers Center (PHY-2317110), the DOE Office of High Energy Physics (DE-SC0018407), and the Air Force Office of Scientific Research (FA9550-19-1-0360). We acknowledge the use of IBM Quantum services for this work. The views expressed are those of the authors, and do not reflect the official policy or position of IBM or the IBM Quantum team.
\end{acknowledgments}

\appendix


\section{A comparison with other approaches}
\label{app:comp}
Various approaches to ``entanglement in time,'' ``entanglement entropy in time'' and ``density matrices for time-separated regions'' have been described in previous work. In this appendix we briefly outline how these proposals differ from ours.

We find it useful to depict various tensor objects diagrammatically. Some existing proposals assign a density matrix $X$ to time-separated regions. In general, it is a complicated tensor object. For the purpose of defining the hermitian conjugate $X^\dagger$ and the moments $\Tr X^n$, we may regard it as a linear map from a certain Hilbert space $\Hc_{in}$ to $\Hc_{out}$: 
\beq
X: \Hc_{in} \ra \Hc_{out}.
\eeq
In turn, $\Hc_{in/out}$ is a tensor-product of various physical Hilbert spaces (or their complex conjugates), possibly located on different time slices.
Our diagrammatic notation uses filled circles for ``out'' and empty circles for ``in''. Colors distinguish different tensor factors; an ``in'' factor is mapped to a particular ``out'' factor if the corresponding filled and open circles are the same color.

Using this notation,  the superdensity matrix $S$ defined in \cite{Cotler:2017anu} for two time-separated regions $A$ (red) and $B$ (green) has the representation
\beq
\label{eq:super}
    \includegraphics{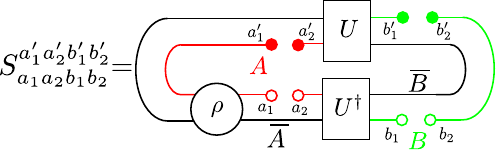}
\eeq

\noindent Mathematically, $S$ is a map
\beq
S: \Hc_A \otimes \Hc^*_A \otimes 
\Hc_B \otimes \Hc^*_B \ra 
\Hc_A \otimes \Hc^*_A \otimes 
\Hc_B \otimes \Hc^*_B.
\eeq
Unlike our construction of $T$ in eq.~(\ref{eq:TAB}), which provided slots on only the forward (i.e., lower) branch of the tensor diagram, $S$ has slots on both the forward and backward (i.e., upper) branch. Furthermore, for $S$ all input indices are on the lower branch and all output indices are on the upper branch.

Because of this assignment of indices, in $\Tr S^n$ the upper branch of each copy of $S$ is contracted with the lower branch of the next copy. Due to this connection of the upper and lower branches, the moments of $S$, unlike the moments of $T$, are not simply related to the R\'enyi (and Tsallis) entropies evaluated in RQFT. In addition, measurement of $\Tr S^n$, in contrast to the measurement protocol for $\Tr T^n$ described in Sec.~\ref{sec:measure}, requires extensive use of axillary systems.
We note that, as explained in \cite{Cotler:2017anu}, other approaches such as consistent histories \cite{griffiths1984consistent} and the multi-state formalism \cite{Aharonov_2009}  can be obtained from $S$. The pseudodensity matrix of \cite{Fitzsimons:2013gga} can also be obtained from $S$ by appropriately projecting it onto single-qubit measurements. 

Another approach is based on the Feynman-Vernon influence functional \cite{Feynman:1963fq,sonner2021influence}. When used to characterize the time correlations of a fixed subsystem $A$, the influence functional for $A$ is obtained by ``integrating out'' the complementary subsystem $\bar A$, which is regarded as the environment of $A$. By vectorizing the influence functional, and also allowing the subsystem $B$ at time $t$ to be distinct from the subsystem $A$ at time zero, we obtain the tensor network representation
\beq
\label{eq:vec_inf}
\includegraphics{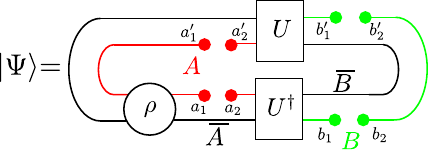}
\eeq
Because operator insertions are permitted on both the forward and backward branches in the Keldysh description of time evolution, there are slots for contracting operators on both branches for both subsystems $A$ and $B$, and all slots are output spaces because the influence functional has been vectorized.
The conventional entanglement entropy evaluated for the state vector $| \Psi \ket$ is called ``temporal entanglement'' \cite{Ba_uls_2009,sonner2021influence,Lerose:2021svg,Luchnikov:2024qvl,Thoenniss:2024nlw,Carignano:2023xbz,Carignano:2024jxb,Bou-Comas:2024pxf}. In general, it too is not related in any obvious way to RQFT computations, although one exception is when this approach is applied to the amplitude $\bra \psi_0 | e^{-i H t} | \psi_0 \ket$ (rather than a correlation function in (\ref{eq:vec_inf})) for an evolving wavefunction to return to its initial state, which can be conveniently computed in RQFT \cite{Carignano:2024jxb}. In addition, $|\Psi \ket$ defined this way does not have unit norm, and has to be normalized by hand.

Our construction is more closely related to the ``ideal two-point quantum correlator''
\cite{Buscemi:2013xlk} and "state-over-time" formalism \cite{Leifer_2013}. 
In  our diagrammatic notation, the ideal two-point quantum correlator $\Tc$ is
\beq
\includegraphics{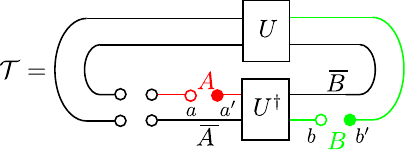}
\eeq
As is evident from this definition, $\Tc$ does not depend on the density matrix $\rho$; instead, $\rho$ should be provided as an extra input. Thus $\Tc$ also includes the parts of the Hilbert space $\bar{A}, \bar{B}$ which are complimentary to $A,B$, making the construction more involved. The image $\Tc(\rho)$, resulting from contracting this diagram with $\rho$, equals our proposed $T_{AB}$.

There are many different constructions of states-over-time \cite{Leifer_2013,Sutter_2016,Fullwood:2022rjd,Parzygnat:2022pax,Parzygnat:2023hcc,Parzygnat_2023}. One variant of it resembles $T_{AB}$ except that it breaks the evolution diagram on both the forward and backward contour:
\beq
\includegraphics{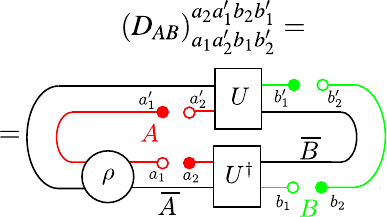}
\eeq
making it a map
\beq
D_{AB}: \Hc_A \otimes \Hc_A \otimes 
\Hc_B \otimes \Hc_B \ra 
\Hc_A \otimes \Hc_A \otimes 
\Hc_B \otimes \Hc_B.
\eeq
An equivalent RQFT path-integral can be represented as
\beq
\includegraphics{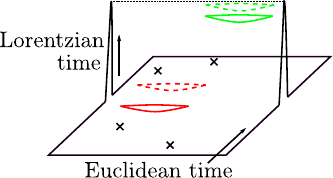}
\eeq
where the extra $\Hc_A \otimes \Hc_B$ factors are 
represented by the dashed lines to emphasize that they lie on the other side of the Lorentzian evolution fold.

Our $T_{AB}$ can be obtained by tracing out these extra $\Hc_A \otimes \Hc_B$ factors. 
It is this tracing out operation that makes $\Tr T_{AB}^n$ well-defined in RQFT, unlike $\Tr D_{AB}^n$. Moments $\Tr D_{AB}^n$ can in principle be computed but they are badly UV divergent because the two copies of $\Hc_A$ and $\Hc_B$ are only separated by the infinitesimal $i \varepsilon$. Apart from the standard UV divergence coming from the interval's endpoints, which can be easily subtracted, there is an additional divergence because the two intervals are close together, and it is not obvious how to subtract this additional divergence.

Finally, we note that Refs. \cite{Horsman_2017,Fullwood:2022rjd,Lie:2023ahy} discuss axiomatic approaches to temporal generalizations of density matrices.

\section{Spectrum of $T_{AB}$ and $T_{\bar{A}^t B}$}
\label{app:spectrum}
To see that $T_{AB}$ has the same spectrum as $T_{\bar{A}^tB}$ we show that all their moments agree: $\Tr T_{AB}^n =\Tr T_{\bar{A}^t B}^n$ for each $n$. We observe that in the tensor diagram representing $\Tr T_{AB}^n$, the copies of $A$ are cyclically connected, as are the copies of $B$.
For example, $T_{AB}^3$ ($A$ is red, $B$ is green) can be expressed as
\begin{center}
\includegraphics{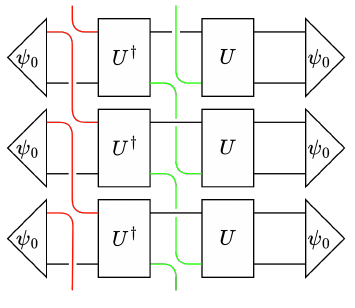}
\end{center}
and when taking the trace the red wires at top and bottom are contracted, as are the two green wires. By shuffling the copies of the bra $\langle \psi_0|$ downward on the left side, we obtain the diagram
\begin{center}
\includegraphics{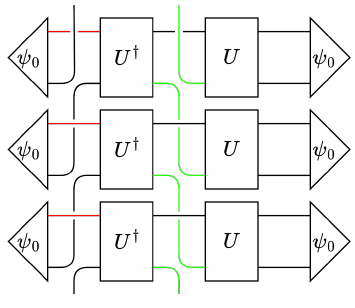}
\end{center}
where the contraction of wires at the top and bottom is now understood.  This is a diagrammatic representation of $\Tr T_{\bar{A}^t B}^3$. The same argument applies to all the other moments.

\section{Bound on the singular values}
\label{app:bound-values}
In this appendix we prove that $T$ satisfies property 7 stated in Sec.~\ref{sec:properties}. 

Note that the Schatten $\infty$-norm can be expressed as
\beq
\label{eq:infty_norm}
||T||_\infty = \sup_{v,y} |\bra y | T | v \ket|, \ ||v||_2=||y||_2=1.
\eeq
where $v,y$ are vectors and 
$||v||_2$ is the standard Euclidean 2-norm.

In our case, the input (or output) space of $T$ is the tensor product of $\Hc_A \otimes \Hc_B$.
So the vector $v$ is a vector in $\Hc_A \otimes \Hc_B$. Therefore, we can use the Schmidt decomposition to write this vector as
\beq
| v \ket = \sum_{\alpha} v_\alpha  | b_\alpha \ket | a_\alpha \ket,
\eeq
where $\{| a_\alpha \ket\}_{\alpha}$ and $\{| b_\alpha \ket\}_{\alpha}$ are orthonormal sets. In this basis, the Euclidean 2-norm is
\beq
(||v||_2)^2 = \sum_\alpha |v_\alpha|^2.
\eeq
Similarly, we can decompose $| y\ket $ as:
\beq
| y \ket = \sum_{\beta} y_\beta  | \tilde{b}_\beta \ket | \tilde{a}_\beta \ket.
\eeq
Then, the corresponding matrix element looks like a sum over correlators:
\begin{align}
&\bra y | T | v \ket = \\ 
& =\sum_{\alpha \beta} v_\alpha y_\beta^* \Tr \Bigg[
\rho \bigg( | a_\alpha \ket \bra \tilde{a}_\beta | \otimes \id_{\bar{A}} \bigg)   U^\dagger  \bigg( |  b_\alpha \ket \bra \tilde{b}_\beta | \otimes \id_{\bar{B}} \bigg)  U 
\Bigg]. \nonumber
\end{align}
We can think of $| a_\alpha \ket \bra \tilde{a}_\beta | \otimes \id_{\bar{A}}$ and $|  b_\alpha \ket \bra \tilde{b}_\beta | \otimes \id_{\bar{B}}$ as operators $\Oc_A$ and $\Oc_B$. Now, we will use Cauchy-Schwartz type inequality to bound the absolute value of the trace,
\begin{equation}
 |\text{Tr} (X^{\dagger} Y) |^2 \leq \text{Tr} (X^{\dagger} X) \text{ Tr} ( Y^{\dagger} Y).
\label{eq:cauchy}
\end{equation}

Note that $\rho$ is positive semi-definite, so $\rho = M^{\dag}M$ for some matrix $M$. Using the cyclic property of the trace, 

\begin{align}
&\Tr \left( \rho \Oc_A U^{\dag} \Oc_B U  \right) = \Tr \left(M^{\dag} M  \Oc_A U^{\dag} \Oc_B U \right) = \\ \notag
& \Tr \left(M  \Oc_A U^{\dag} \Oc_B U M^{\dag} \right)
\end{align}
Then, using  the inequality (\ref{eq:cauchy}), 
\begin{align}
\label{eq:oper_CS}
&|\Tr \left(M  \Oc_A U^{\dag} \Oc_B U M^{\dag} \right)| \leq \\ \notag
& \leq \left[\Tr \left(M  \Oc_A U^{\dag} U \Oc_A^{\dag} M^{\dag}\right) \Tr \left( M U^{\dag} \Oc_B^{\dag} \Oc_B U M^{\dag} \right)\right]^{\frac{1}{2}}
\end{align}
Note that
\begin{align}
\Oc_A \Oc_A^{\dag} &= \left(| a_\alpha \ket \bra \tilde{a}_\beta | \otimes \id_{\bar{A}}\right) \left( |\tilde{a}_\beta \ket \bra  a_\alpha | \otimes \id_{\bar{A}} \right)\\
&=  | a_\alpha \ket  \bra  a_\alpha | \otimes \id_{\bar{A}}
\end{align}
and
\begin{align}
\Oc_B^{\dag} \Oc_B &= \left(| \tilde{b}_\beta \ket \bra b_{\alpha} | \otimes \id_{\bar{B}} \right) \left( |  b_\alpha \ket \bra \tilde{b}_\beta | \otimes \id_{\bar{B}}\right) \\
&= | \tilde{b}_\beta \ket  \bra  \tilde{b}_\beta | \otimes \id_{\bar{B}}.
\end{align}
Since $U$ is a unitary operator, $UU^{\dag} = U^{\dag}U = \id$, we can use the cyclic property of the trace to reduce the first factor inside the square root to
\beq
\Tr \left( M^{\dag} M \Oc_A \Oc_A^{\dag}  \right) = \Tr \left( \rho | a_\alpha \ket  \bra  a_\alpha | \otimes \id_{\bar{A}} \right) .
\eeq
For the second factor in the square root, we have
\beq
\Tr \left(M^{\dag} M U^{\dag} | \tilde{b}_\beta \ket  \bra  \tilde{b}_\beta | \otimes \id_{\bar{B}} U  \right) = \Tr \left(U \rho U^{\dag} | \tilde{b}_\beta \ket  \bra  \tilde{b}_\beta | \otimes \id_{\bar{B}}   \right).
\eeq

Therefore, combining the above results, 
\begin{widetext}
\begin{align}
&||T||_\infty = \sup_{v,y} |\bra y | T | v \ket| = \\ \nonumber
& = \sup_{v,y} \Bigg| \sum_{\alpha \beta} v_\alpha y_\beta^* \Tr \Bigg[\rho \bigg( | a_\alpha \ket \bra \tilde{a}_\beta | \otimes \id_{\bar{A}} \bigg) U^\dagger  \bigg( |  b_\alpha \ket \bra \tilde{b}_\beta | \otimes \id_{\bar{B}} \bigg)  U 
\Bigg] \Bigg| \leq  
\notag \\ 
& \leq \sup_{v,y} \sum_{\alpha \beta} \Bigg| v_\alpha y_\beta^* \Tr \Bigg[
\rho \bigg( | a_\alpha \ket \bra \tilde{a}_\beta | \otimes \id_{\bar{A}} \bigg)  U^\dagger  \bigg( |  b_\alpha \ket \bra \tilde{b}_\beta | \otimes \id_{\bar{B}} \bigg)  U 
\Bigg] \Bigg| \leq  \notag \\
& \leq \sup_{v,y} \sum_{\alpha \beta} |v_\alpha y_\beta^*| \Biggl( \Tr\Biggl[\rho | a_\alpha \ket  \bra  a_\alpha | \otimes \id_{\bar{A}}\Biggr]  \Tr \Biggl[U \rho  U^{\dag} | \tilde{b}_\beta \ket  \bra  \tilde{b}_\beta | \otimes \id_{\bar{B}}   \Biggr] \Biggr)^{\frac{1}{2}} = \nonumber \\
& = \sup_{v,y} \left( \sum_{\alpha} |v_{\alpha}| \left[ \bra a_{\alpha} | \rho_A | a_{\alpha} \ket\right]^{\frac{1}{2}}  \right) \left(\sum_{\beta} |y_{\beta}^*| \left[ \bra \tilde{b_{\beta}} | \rho_B | \tilde{b_{\beta}} \ket \right]^{\frac{1}{2}} \right). \nonumber
 \end{align}
 \end{widetext}
 Applying Cauchy--Schwarz to both sums,
 \begin{align}
 ||T||_{\infty} 
  \leq \left[\Tr(\rho_A) \right]^{\frac{1}{2}} \left[\Tr(\rho_B) \right]^{\frac{1}{2}} = 1.
\end{align}

\section{Proof of Theorem \ref{th:mono}}
\label{app:mono}
To prove Theorem \ref{th:mono}, we use the 
dual property of Schatten norms: 
\beq
||\rho_B||_p = \max_{||Y_B||_q=1} |\Tr( \rho_B Y_B)| , \ 1/p+1/q=1,
\eeq
where the maximum is taken over operators $Y_B$ with unit $||Y_B||_q$ norm, acting on $\Hc_B$. The maximum  is attained at some $Y^*_{B}$ because $|\Tr( \rho_B Y_B)|$ is continuous and $||Y_B||_q=1$ ranges over a compact space. 
In turn, $\Tr(Y_B^* \rho_B)$ can be written in terms of $T_{AB}$:
\beq
\begin{split}
||\rho_B||_p &= |\Tr(Y^*_B \rho_B)| = |\Tr(T_{AB} \id_A \otimes Y_B^*)|   \\
&=|\Tr(T_{AB} | \psi_A \ket \bra \psi_A| \otimes Y^*_B)| \le ||T_{AB}||_p.
\end{split}
\eeq
The second equality holds because $\Tr_A T_{AB} =\rho_B$. The third equality holds because, as specified in the premise of Theorem \ref{th:mono}, the initial state is $\rho = | \psi_A \ket \bra \psi_A| \otimes \rho_{\bar{A}}$; hence $\rho \id_A \otimes Y_B^*= \rho | \psi_A \ket \bra \psi_A|\otimes Y_B^*$. The third equality is illustrated by the following diagram:
\beq
\includegraphics{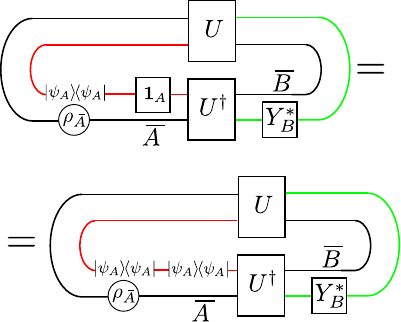}
\eeq
The last inequality is H\"older's inequality plus the trivial statement that $|| \ | \psi_A \ket \bra \psi_A| \ ||_q=1$. 

\section{Bounds on commutators}
\label{app:bounds-commutators}
To prove Theorems \ref{th2} and \ref{th3}, we repeatedly use H\"older's inequality and keep in mind that it provides a tight bound. The commutator can be rewritten as
\beq
i \bra [\Oc_A(0), \Oc_B(t)] \ket = \Tr( M \Oc_A \otimes \Oc_B) = \Tr(M_A \Oc_B),
\eeq
where $M=i(T_{AB}-T_{AB}^\dagger)$  and
the operator $M_A$ is defined by simply contracting $\Oc_A$ with $M$:
\beq
    \includegraphics[width=0.7\linewidth]{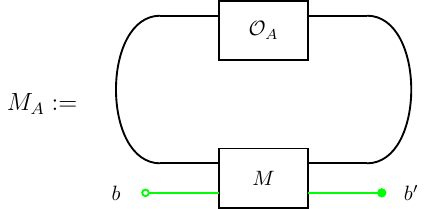}
\label{fig:MA_def}
\eeq
Using H\"older's inequality
\beq
|\Tr(\Oc_B M_A)| \leq
||\Oc_B||_2 \sqrt{\Tr(M_A^{\dag}M_A)}.
\eeq
Now we rotate the tensor diagram for $\Tr(M_A^{\dag}M_A)$ by 90 degrees and treat it as a matrix element of $M_T M_T^\dagger$ (defined by the eq.~(\ref{eq:MT})) evaluated on the state $| \Oc_A )$ (the vectorized $\Oc_A)$:
\begin{align}
\Tr(M_A^{\dag}M_A) &= ( \Oc_A | M_T M_T^{\dag} | \Oc_A) \\
&\leq ||\Oc_A||_2^2 ~ \max \lambda(M_T M_T^{\dag}),
\end{align}
where $\max \lambda(M_T M_T^{\dag})$ is the maximum eigenvalue of $M_T M_T^{\dag}$.
Therefore,
\beq
\label{eq:maxl_ineq}
| \langle \left[\Oc_A(0), \Oc_B(t) \right]  \rangle | \leq ||\Oc_A||_2 ||\Oc_B||_2 \|M_T\|_\infty.
\eeq
Also note that each inequality we used was tight, hence there exist operators $\Oc_A, \Oc_B$ for which (\ref{eq:maxl_ineq}) is saturated.
This proves \textbf{Theorem} \ref{th3} in the main text.

To prove \textbf{Theorem} \ref{th2}, we note that 
$M_T M_T^\dagger$ is positive-definite map  from $\Hc_B \otimes \Hc_B^*$ onto itself.
Furthermore, for positive-definite matrices the maximum eigenvalue is greater than or equal to the sum of all eigenvalues divided by the dimension:
\beq
\begin{split}
\sqrt{\max \lambda(M_T M_T^{\dag})} \geq \frac{\sqrt{\Tr(M_T M_T^{\dag})}}{\dim \Hc_B} = \\ = \frac{\sqrt{\Tr(M^2)}}{\dim \Hc_B} 
=\frac{||T-T^{\dag}||_2}{\dim \Hc_B}.
\end{split}
\label{lg7}
\eeq
Here we used the property $\Tr M_T M_T^{\dag} = \Tr M^2$, which follows immediately from the definition of $M_T$ as a realignment of $M$, eq.~(\ref{eq:MT}).

The same bound can be found with $\dim \Hc_B$ replaced by $\dim \Hc_A$ because an analogous derivation applies to the operator $M_B$ instead of $M_A$. Therefore we may choose the denominator to be $\min (\dim \Hc_A,\dim  \Hc_B)$ to obtain a tighter bound.

\section{Details on free-fermion lattice computation}
\label{app:free-fermion}
Given that all spacelike and timelike correlation functions obey Wick's theorem, $T$ must have Gaussian form, including correlations between fermions $\psi_{A/B}$ located on $A/B$ respectively:
\beq
\begin{split}
T = \Nc^{-1} \exp \Bigg( -\sum_{s_1 s_2 \in A} a_{s_1 s_2} \bar{\psi_A}(s_1)  \psi_A(s_2) - \\
- \sum_{s_1 s_2 \in B} b_{s_1 s_2} \bar{\psi_B}(s_1) \psi_B(s_2)- \\  - \sum_{ s_1 \in A, s_2 \in B}c_{s_1 s_2} \bar{\psi_A}(s_1) \psi_B(s_2) - \\ -\sum_{s_1 \in A, s_2 \in B} d_{s_1 s_2} \bar{\psi_B}(s_2) \psi_A(s_1)   \Bigg),
\end{split}
\eeq
where $a_{s_1 s_2},b_{s_1 s_2},c_{s_1 s_2},d_{s_1 s_2}$ are complex numbers (not operators) and $\Nc$ is the normalization factor. For convenience, we can group $\psi_{A/B}$ into a single vector $\Psi= (\psi_A, \psi_B)^T$ and $a,b,c,d$ into a block matrix $t_{s_1 s_2}$ to rewrite the above equation as
\beq
T = \Nc^{-1} \exp \l( -\sum_{s_1 s_2} t_{s_1 s_2} \bar{\Psi}(s_1) \Psi(s_2) \r)
\eeq
For the free-fermion computation in Sec.~\ref{sec:free-fermions}, in order to find $t_{s_1 s_2}$, and hence the eigenvalues of $T$, one forms a correlation matrix consisting of 4 blocks:
\beq
C = \begin{pmatrix}
\bra \bar{\psi}(0,s) \psi(0,y) \ket & -\bra  \psi(t,y) \bar{\psi}(0,s)  \ket \\
\bra \bar{\psi}(t,s) \psi(0,y) \ket & \bra \bar{\psi}(t,s) \psi(t,y) \ket
\end{pmatrix}, \ 
\eeq
where $s \in A$ and $y \in B$.
An additional minus sign in one of the elements arises because $T$ corresponds to a \textit{fixed} operator ordering --- $\psi(t,y)$ has to be before $\bar \psi(0,s)$.

At half-filling the exact correlators are:
\beq
\begin{split}
&\bra \bar{\psi}(t,s)  \psi(0,0) \ket\\
&= 
\frac{1}{4} \Bigg( - i t f(1,(3-s)/2,(3+s)/2,-t^2/4) +  \\
& +2 f(1,(2-s)/2, (2+s)/2,-t^2/4) \Bigg),
\end{split}
\eeq
where 
\beq
f(a,b_1,b_2,y) = \frac{\phantom{f}_1 F_2(a,b_1,b_2,y)}{\Gamma(b_1) \Gamma(b_2)}
\eeq
and $\bra \psi(t,s) \bar{\psi}(0,0) \ket = e^{i \pi s} \bra \bar{\psi}(t,s) \psi(0,0) \ket$.
Then the single-particle entanglement Hamiltonian $t_{s_1 s_2}$ can be found \cite{Peschel_2003} through the standard formula $t^T=\log((1-C)/C)$.

This procedure can also be applied to $T^\dagger$. As we have explained in the main text, in general $[T,T^\dagger] \neq 0$, which is also the case here.
This is why in computing $\Tr(T T^\dagger)$ one cannot sum the absolute values of the eigenvalues of $T$. Instead, one has to multiply $T$ and $T^\dagger$, which can be easily done because both are Gaussian. But, unfortunately, $\Tr(T T^\dagger)$ is not always well-defined in the continuum limit as we will explain in Appendix \ref{app:ttbar}.

\section{CFT 2-replica computation}
\label{app:cft2}
In $(1{+}1)$-dimensional CFTs the torus partition function $Z_2(\tau, \bar{\tau)}$ depends on the complex torus modulus $\tau$ and its complex conjugate.
The purity $\Tr \rho_{AB}^2$ for two intervals $A=[z_1,z_2],B=[z_3,z_4]$ can be mapped to the torus partition function \cite{Lunin:2000yv}:
\beq
\begin{aligned}
& \Tr \rho_{AB}^2 = Z_2(\tau,\bar{\tau}) 2^{-2c/3}  \\
&\times \l( \frac{\ep^4}{(z_2-z_1) (\bar{z}_2-\bar{z}_1) (z_4-z_3) (\bar{z}_4-\bar{z}_3) } \r)^{1/8}  \\
&\times \l( \frac{x}{(1-x)^2} \r)^{-c/24}
\l( \frac{\bar{x}}{(1-\bar{x})^2} \r)^{-c/24},
\end{aligned}
\eeq
 The cross-ratio $x$ and modular parameter $\tau$ are given by
\beq
x= \frac{\theta_2(e^{2 \pi i \tau})^4}{\theta_3(e^{2 \pi i \tau})^4}, \ \tau = \frac{i}{2} \frac{K(1-x)}{K(x)},
\eeq
\beq
\label{eq:barx_bartau}
\bar{x}= \frac{\theta_2(e^{-2 \pi i \bar{\tau}})^4}{\theta_3(e^{-2 \pi i \bar{\tau}})^4}, \ \bar{\tau} = -\frac{i}{2} \frac{K(1-\bar{x})}{K(\bar{x})}.
\eeq
For intervals located at the same time-slice, $x \in (0,1)$ and $\tau$ is purely imaginary. We analytically continue away from this regime, by treating $z, \bar{z}$ as the light-cone coordinates $(u,v)=(s-t,s+t)$. 

One comment concerns modular invariance. $Z_2(\tau,\bar{\tau})$ is invariant under torus modular transformations if $\bar{\tau}$ is the complex conjugate of $\tau$. Under the analytic continuation,
the cross-ratios $x, \bar{x}$ remain real (except the small imaginary parts responsible for operator ordering, as explained in Sections \ref{sec:def}, \ref{sec:free-fermions}). However, $\bar{\tau}$ is no longer the complex conjugate of $\tau$ as follows from eq.~(\ref{eq:barx_bartau}); therefore $Z_2$ is not invariant under the T-modular transformation $\tau \ra \tau+1$. However, invariance under the S transformation $\tau \ra -1/\tau$ remains unbroken. To put it another way, we are performing the analytic continuation from $\mathbb{R} \ni -i \tau > 0 $. The S-modular transformation preserves this domain, which is why it remains unbroken, unlike the T-modular transformation. 

We note in passing that the procedure outlined above can be used to analytically continue $\Tr \rho_{AB}^2$ to the regime $\mathbb{R} \ni x=\bar{x}<0$. In fact there are time separations that lead to exactly this regime. Continuation of $\Tr \rho_{AB}^2$ to $\mathbb{R} \ni x=\bar{x}<0$ is also needed for the computation of entanglement negativity \cite{Calabrese:2012nk}. However, the two continuations are generally different. For example, for the negativity computation the continuation into the $x<0$ domain ensures that the answer is real and preserves the T-modular transformation. 

For holographic theories the torus partition function can be computed by the saddle-point approximation, and the dominant bulk geometry depends on the value of $\tau$ \cite{Dijkgraaf:2000fq}. However, given that we are performing an analytic continuation it is not obvious which geometry (saddle point) will dominate. This would be interesting to investigate.

\section{$\Tr T^2$ and $\Tr T T^\dagger$ as correlators}
\label{app:ttbar}

In this Appendix we explain why $\Tr(T T^\dagger)$ can be very sensitive to lattice effects and might not behave smoothly in the continuum limit.
A SWAP-operator can be rewritten as a sum over a complete set $\{ \Oc_\alpha \}$ of properly normalized hermitian operators:
\beq
\text{SWAP}_{AA'} = \sum_\alpha \Oc_{A,\alpha} \otimes \Oc_{A',\alpha}.
\eeq
Using that, one can rewrite $\Tr T^2$ and $\Tr T T^\dagger$ as sums of products of correlators:
\begin{align}
&\Tr T^2 \nonumber\\
&= \Tr \text{SWAP}_{AA'}\text{SWAP}_{BB'}\left(T_{AB}\otimes T_{A'B'}\right)\nonumber\\
&= \Tr\sum_{\alpha,\beta}\left(\Oc_{A,\alpha}\otimes\Oc_{A',\alpha}\right)
\left(\Oc_{B,\beta}\otimes\Oc_{B',\beta}\right)\left(T_{AB}\otimes T_{A'B'}\right) \nonumber\\
&= \sum_{\alpha,\beta} \Tr\left(\Oc_{A,\alpha}\Oc_{B,\beta}T_{AB}\right)\Tr\left(\Oc_{A',\alpha}\Oc_{B',\beta}T_{A'B'}\right)\nonumber\\
&=\sum_{\alpha,\beta} \bra \Oc_{A,\alpha}(0) \Oc_{B,\beta}(t)  \ket^2,
\end{align}
and similarly
\begin{align}\label{eq:TTdagger-correlator}
\Tr T T^\dagger 
&= \sum_{\alpha,\beta} \bra \Oc_{A,\alpha}(0) \Oc_{B,\beta}(t)  \ket \bra \Oc_{B,\beta}(t)\Oc_{A,\alpha}(0)   \ket\nonumber\\
&= \sum_{\alpha,\beta} |\bra \Oc_{A,\alpha}(0) \Oc_{B,\beta}(t)  \ket|^2.
\end{align}
The trouble is that lattice effects might accumulate in $\Tr (T T^\dagger)$ because of the absolute value in eq.~\ref{eq:TTdagger-correlator}.

\begin{figure}
    \centering
    \includegraphics[scale=1.7]{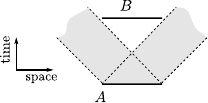}
    \caption{In any CFT, operators on $A$ can have nonzero (anti-)commutators only with operators supported in the shaded area, because all excitations propagate strictly at the speed of light.
    Region $B$ is time separated from $A$, but operators in $A$ cannot influence operators in $B$; hence $T_{AB}=T^\dagger_{AB}$.}
    \label{fig:pure_time}
\end{figure}

Correlators $\bra \Oc_{A,\alpha}(0) \Oc_{B,\beta}(  t)  \ket$ can behave quite differently on the lattice than in the continuum. For example, the lattice answer can contain additional highly oscillatory terms that do not occur in the continuum, as happens for the random hopping model (\ref{eq:RH}). In the continuum, this model becomes relativistically invariant and moreover all excitations are massless and propagate at the speed of light. Therefore, all commutators of bosonic operators (or anti-commutators of fermionic operators) are nonzero only on the light cone. This is illustrated by Figure \ref{fig:pure_time}, where bosonic (fermionic) operators on $A$ have nonzero (anti-)commutators only with operators in the shaded area.

Because commutators $\bra [\Oc_{A,\alpha}(0), \Oc_{B,\beta}(t)]\ket =
2 i \Im \bra \Oc_{A,\alpha}(0) \Oc_{B,\beta}(t) \ket$ vanish, we conclude that 
$T_{AB}=T_{AB}^\dagger$ in the continuum.
However, on the lattice these imaginary parts are not zero; instead they are highly oscillatory.
These oscillations cancel out in $\Tr T^2$, but may not cancel out in $\Tr T T^\dagger$ because of the absolute value; therefore  $\Tr T T^\dagger$ can be very large on the lattice even though it vanishes in the continuum. For example, it is easy to compute $\Tr(T T^\dagger)$ numerically for the random hopping model (\ref{eq:RH}) in the kinematical regime shown in Figure \ref{fig:pure_time}, as outlined in Appendix \ref{app:free-fermion}.
The answer for $\Tr(T T^\dagger)$ is significantly larger than $\Tr T^2$ and it diverges exponentially $\sim e^{1/a}$ when the lattice spacing $a$ goes to zero.

\section{Free fermion CFT results}
\label{app:free-cft}

For the CFT computation in Sec.~\ref{sec:cft}, we need two identities for theta-functions and the Dedekind $\eta$-function  (again, we are using Wolfram Mathematica notation):
\beq
\theta_3^4(e^{2 \pi i \tau}) - 
\theta_2^4(e^{2 \pi i \tau}) = \theta_4^4(e^{2 \pi i \tau}),
\eeq
and 
\beq
2 \eta^3(2 \tau) = \theta_2(e^{2 \pi i \tau}) \theta_3(e^{2 \pi i \tau})
\theta_4(e^{2 \pi i \tau}).
\eeq
This is enough to show that eqns. (\ref{eq:cft_tt}) and (\ref{eq:ff_z}) together
reproduce eq.(\ref{eq:two_int_renyi}).

\section{OPE limits}
\label{app:ope}
Twist operators are not, strictly speaking, local: they are end points of a topological defect line which connects different replica copies. Each defect line has one twist and one anti-twist operator. For pure states we can draw the defect lines for two intervals in two distinct ways:
\begin{center}
\includegraphics[scale=1.4]{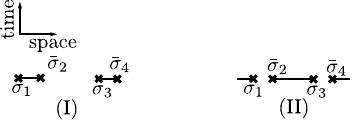}
\end{center}
When $\bar{\sigma}_2$ is close to $\sigma_3$, it is convenient to use the second way to illustrate that $\sigma_3$ and $\bar{\sigma}_2$, along with their defect line, become close and separated from the rest of the operators. This applies for both the standard entanglement entropy when the intervals are on the same time-slice:
\begin{center}
\includegraphics[scale=1.4]{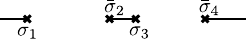}
\end{center}
and for our entanglement in time, when they are on different time-slices but end-points of the intervals are close to null-separation (dashed lines indicate the light cone):
\begin{center}
\includegraphics[scale=1.4]{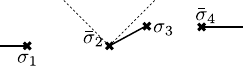}
\end{center}

However, in the Lorentzian setting the OPE limit is more subtle \cite{mack1977convergence,Pappadopulo:2012jk}: one has to make sure 
the two operators that become close are away from the past and the future light cones of other operators.\footnote{We thank David Simmons-Duffin for an illuminating discussion on this point.}
When $A,B$ are on top of each other and $\sigma_1$ is very close to $\sigma_3$, their defect lines still lie in the future light cone of $\bar{\sigma}_2$ and $\bar{\sigma}_4$ (dashed lines indicate the light cone):
\begin{center}
\includegraphics[scale=1.4]{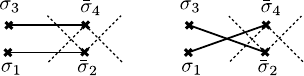}
\end{center}
hence one cannot simply take the standard OPE limit of $\sigma_1 \sigma_3$. This can be illustrated by the free fermion answer (\ref{eq:two_int_renyi}) where the limit $z_1 \ra z_3$ is not reproduced by the standard $\sigma \sigma$ OPE. Again, this is different from the case of negativity where the two twists $\sigma \sigma$
lie on the same time-slice and there are no obstacles to taking their OPE \cite{Calabrese:2012nk}.

\bibliography{refs}
\end{document}